\pdfoutput=1
\documentclass[aps,pre,onecolumn,amsmath,amssymb,nofootinbib,superscriptaddress,showpacs,longbibliography]{revtex4-2}
\usepackage{graphicx}
\usepackage{physics}
\usepackage{bbold}
\usepackage{hyperref}
\usepackage{natbib}
\usepackage{latexsym}
\usepackage{amsmath, amssymb, amsthm}
\usepackage{mathtools}
\usepackage[dvipsnames]{xcolor}
\usepackage{comment}
\usepackage{wasysym}
\usepackage[bottom]{footmisc}
\usepackage{float}
\usepackage{appendix}
\usepackage[T1]{fontenc}
\usepackage[utf8]{inputenc}
\usepackage[english]{babel} 

\newcommand{\kmin}{k_{\textrm{min}}}
\newcommand{\kmax}{k_{\textrm{max}}}
\newcommand{\KMAX}{K_{\textrm{max}}}

\newcommand{\rRR}{r_{\textrm{RR}}}
\newcommand{\rER}{r_{\textrm{ER}}}

\newcommand{\change}[1]{}

\usepackage{comment}
\begin{document}

\title{Strongly clustered random graphs via triadic closure: Degree correlations and clustering spectrum}

\author{Lorenzo Cirigliano}
\affiliation{Dipartimento di Fisica Universit\`a \href{https://ror.org/02be6w209}{La Sapienza}, I-00185 Rome, Italy}

\author{Gareth J. Baxter}
\affiliation{Departament of Physics and i3N, University of Aveiro, Campus Universitário de Santiago, 3810-193 Aveiro, Portugal}

\author{G\'abor Tim\'ar}
\affiliation{School of Mathematics, University of Leeds, Leeds LS2 9JT, United Kingdom}

\date{\today}

\begin{abstract}

Real-world networks often exhibit strong transitivity with nontrivial local clustering spectra and degree correlations. Such features are not easily modeled in tractable network models, creating an obstacle to the theoretical understanding of such complex network structures.
Here, we address this problem using a model for strongly clustered random graphs in which each triad of a random network backbone is closed with a certain probability.
Despite the intricate loopy local structure of the graphs obtained, we provide exact expressions for the local clustering spectrum and the degree correlations, filling the gap in the theoretical description of this model for random graphs.
In particular, we find positive degree assortativity accompanying high transitivity, and nontrivial structure in the clustering spectrum. Exact asymptotic analytical results, obtained for uncorrelated locally tree-like backbones, are complemented with extensive numerical characterization of finite-size effects.
\end{abstract}


\maketitle

\vspace{-3cm}


	
	
	\section{Introduction}

	Local features like clustering and neighbor degree correlations are abundant in real-world complex networks and can have a significant effect on network function \cite{newman2018networks}. However, these features are poorly accounted for in established theoretical models of random graphs.
	More specifically,  degree--degree correlations are positively associated with transitivity \cite{newman2002assortative, newman2003social, foster2011clustering, filho2020transitivity}. Local clustering is not typically uniform in nodes of different degrees, as one might expect in a purely random network, but instead the clustering spectrum is usually nontrivial, often decaying with increasing degree  \cite{vazquez2002large, ravasz2003hierarchical, serrano2006clustering, stegehuis2017clustering, van2018triadic}.
	
	Several theoretical results have been obtained in synthetic networks, including strongly regular graphs \cite{li2017clustering}, scale-free networks with hidden variables \cite{van2017local}, and configuration model networks \cite{van2018triadic}. However, a theoretical understanding of the clustering spectra of sparse random networks with tunable global transitivity is lacking.
	
	Random networks with strong clustering can be generated in a number of different ways \cite{holme2002growing, angeles2005tuning, toivonen2006model, bianconi2014triadic}. Analytically tractable models typically rely on some sort of tree-like structure of local motifs---such as trees of loops~\cite{newman2009random}, trees of cliques~\cite{gleeson2009bond} or trees of partial cliques and more complex motifs \cite{karrer2010random, mann2021random}---to achieve the desired clustering.
	These methods produce constrained ensembles of networks which do not, however, include overlapping motifs and loops of arbitrary length.
	Qualitatively, these models show that there are many ways to produce the same level of global clustering, and a careful analysis of local clustering is necessary to characterize them.
	
\begin{figure}[H]
		\centering
		\includegraphics[width=0.8\textwidth]{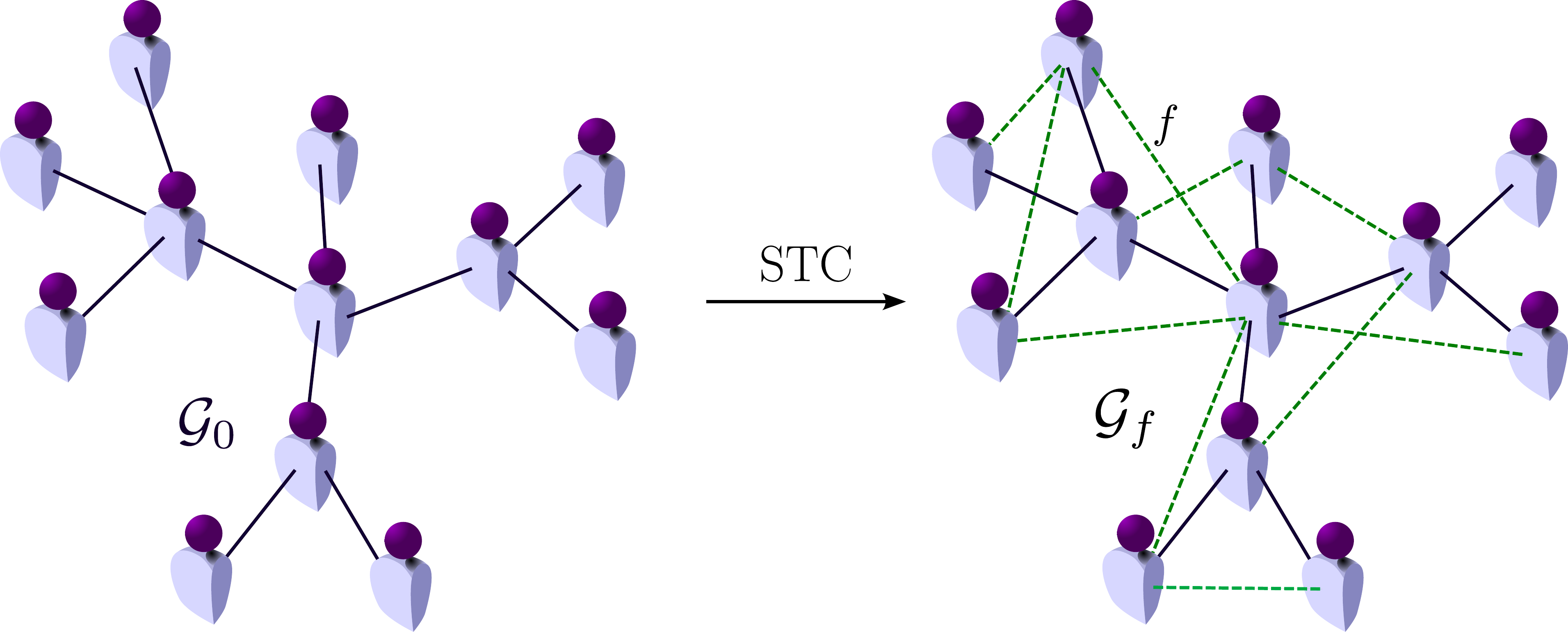}
		\caption{A pictorial representation of the STC mechanism. Imagine some people going to a party. Let us place a link between two individuals if they are friends. This social network \change{before} the party is our backbone $\mathcal{G}_0$ on the left. In our simplifying assumptions, we consider that people initially do not have many friends in common, i.e., $\mathcal{G}_0$ is tree-like. The party begins, and people start talking with their friends. It may happen that two people having a common friend meet and become friends by the end of the party. If that is the case, we put a green dashed link between them. The social network \change{after} the party, $\mathcal{G}_f$, is drawn on the right. Despite the simplicity of the process, the network $\mathcal{G}_f$ clearly exhibits an intricate local structure with overlapping loops.}
		\label{fig:STC_schematic}
	\end{figure}

	The recently introduced Static Triadic Closure (STC) random graph model~\cite{cirigliano2024strongly} overcomes these limitations, efficiently producing highly random graphs with strong clustering which contain overlapping loops and motifs with statistics resembling those of real networks.
	The construction rule for STC random graphs is simple: take a backbone graph $\mathcal{G}_0$, and close each open triad independently with probability $f$ to create a new random graph $\mathcal{G}_f$, see Figure~\ref{fig:STC_schematic}.
	If the starting backbone $\mathcal{G}_0$ is uncorrelated and sparse---implying also local tree-likeness in the infinite-size limit---exact analytical results can be obtained. For instance, site percolation has been exactly solved~\cite{cirigliano2025how} for the case $f=1$, clarifying the role of clustering in determining the universal critical properties of systems defined on complex networks.\change{ STC random graphs with  Erd\H{o}s--R\'enyi backbones were also found to have a surprisingly rich $k$-core structure~\cite{bhat2017exotic}}.
	The analysis of global topological properties, such as the global clustering (transitivity) and small motif densities, of $\mathcal{G}_f$ obtained from uncorrelated random backbones, was developed in~\cite{cirigliano2024strongly}.

	While the STC model is unbiased, it does produce degree correlations, which have not yet been characterized. Furthermore, the understanding of local clustering properties for this model is still lacking. Here, we complete the analysis of STC random graphs, providing a detailed numerical and analytical study of degree correlations and clustering spectrum.

	We show that, by separating edges into those present in the backbone network and those added via triadic closure, and taking advantage of the uncorrelated random nature of the backbone network, a number of exact results can be obtained.
	We give an exact expression for the Pearson correlation coefficient in terms of the moments of the backbone network degree distribution. We find exact formulae for the local clustering spectrum, which in general need to be summed numerically, and we derive the asymptotic behavior.
	
	We find that the STC process always creates assortative correlations, evidenced by a positive Pearson coefficient, and that the effect tends to increase with transitivity. This shows that such correlations are a natural consequence of the unbiased triadic closure process, thus explaining at least some of the assortative correlations found in highly clustered real networks.
	For scale-free backbones we observe sharp transitions in the Pearson correlation coefficient in the infinite-size limit, mirroring the transition observed in the transitivity~\cite{cirigliano2024strongly}.
	We also characterize the local clustering spectrum, observing a qualitative distinction between homogeneous and heterogeneous structures. We characterize the significant finite-size effects present when the backbone network has a power-law \mbox{degree distribution.}
	
	The remainder of this paper is organized as follows. In Section~\ref{sec:STC_definition}, we present some definitions and briefly recall some results on the STC graphs. In Section~\ref{sec:correlations} and Section~\ref{sec:clustering}, we analyze the degree correlations and the local clustering spectrum, respectively, in STC graphs with uncorrelated backbones of arbitrary degree distribution.
	Finally, in Section~\ref{sec:discussion} we summarize our results and discuss their relevance and their implications.

	\section{The Degree Distribution of STC Graphs}
	\label{sec:STC_definition}
	
	In this section we briefly recall some results for the degree distribution of STC random graphs $\mathcal{G}_f$~\cite{cirigliano2024strongly} with an arbitrary uncorrelated random backbone. We also present some novel results for finite-size scale-free STC graphs. We denote by lower case letters quantities related to the backbone network $\mathcal{G}_0$, such as the degree $k$ and the degree distribution $p(k)$, and with capital letters $K$ and $P(K)$ the corresponding quantities for the clustered \mbox{network $\mathcal{G}_f$.}
	
	Let us consider a node with original degree $k$ surrounded by neighbors with original excess degrees $r_1,\dots, r_k$. After the triadic closure, such a node has degree $K=k+\sum_{i=1}^{k}\nu_i$, where $\nu_i$ are random variables representing the number of new neighbors made along the $i$-th original branch. The excess degrees in $\mathcal{G}_0$ are independent random variables with distribution $q(r)=(r+1)p(r+1)/\langle k \rangle$, while the $\nu_i$ are independent binomial random variables with distribution, given $r_i$, $\mathcal{B}(\nu_i|r_i,f)={r_i \choose \nu_i}f^{\nu_i}(1-f)^{r_i-\nu_i}$. The conditional degree distribution is then simply given by $P\left(K|k,\{r_i \}, \{\nu_i \}\right)=\delta\left(K,k+\sum_{i=1}^{k}\nu_i\right)$. Averaging over $P(k,\{r_i \}, \{\nu_i \})=p(k) \prod_{i=1}^{k}q(r_i)\mathcal{B}(\nu_i|r_i,f)$, since $\mathcal{G}_0$ is uncorrelated, we then get
	\begin{equation}
		\label{eq:degree_distribution}
		P(K) = \sum_{k, \{r_i \}, \{\nu_i \}}p(k) \prod_{i=1}^{k}q(r_i)\mathcal{B}(\nu_i|r_i,f) \delta\left(K,k+\sum_{i=1}^{k}\nu_i\right).
	\end{equation}
	\indent Although the r.h.s. of Equation~\eqref{eq:degree_distribution} cannot be computed explicitly, we can easily find an expression for the generating function $G_0(z)=\sum_{K}P(K) z^{K}$ of the degree distribution $P(K)$. Multiplying Equation~\eqref{eq:degree_distribution} by $z^K$ and summing over $K$ we get
	\begin{equation}
		\label{eq:G_0_STC}
		G_0(z)=g_0\left(z g_1(1-f+fz) \right),
	\end{equation}
	where $g_0(z)=\sum_{k}p(k) z^k$ and $g_1(z)=\sum_{r}q(r) z^r$. Equation~\eqref{eq:G_0_STC} will be fundamental in many calculations, as it allows us to express the moments $\langle K^n \rangle$ of the degrees in $\mathcal{G}_f$ in terms of the moments $\langle k^m \rangle$, with $m \leq n+1$, of the degrees in the backbone $\mathcal{G}_0$~\cite{cirigliano2024strongly}. In other words, we can characterize the properties of $\mathcal{G}_f$ from the knowledge of the backbone $\mathcal{G}_0$.
	
	\subsection*{The Case of Power-Law Backbones} 
\begin{figure}[H]
		\centering
		\includegraphics[width=0.98\textwidth]{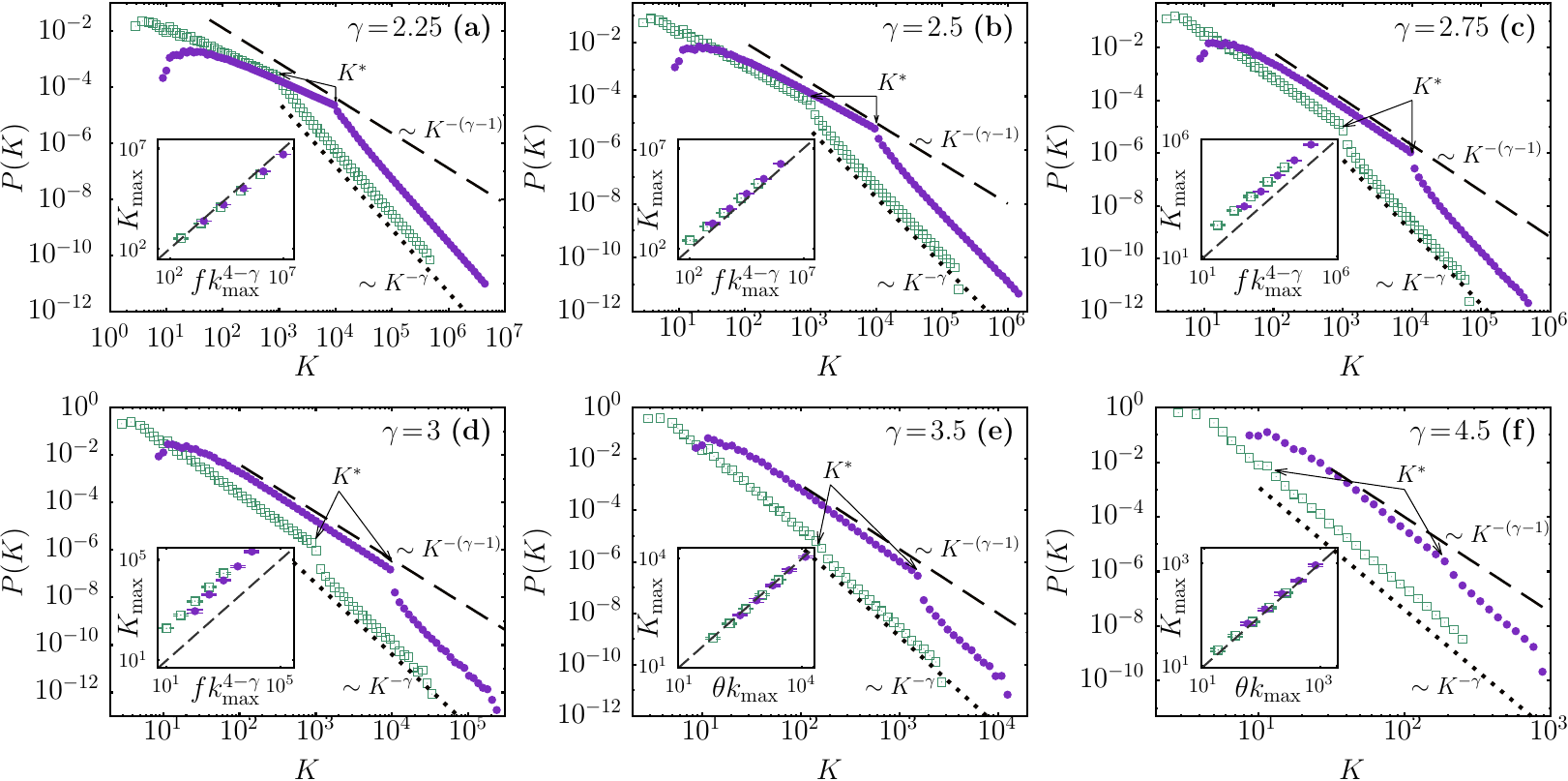}
		\caption{Numerical results for the degree distributions $P(K)$ of STC networks with power-law backbones with exponents (\textbf{a}) $\gamma=2.25$, (\textbf{b}) $\gamma=2.5$, (\textbf{c}) $\gamma=2.75$, (\textbf{d}) $\gamma=3$, (\textbf{e}) $\gamma=3.5$, (\textbf{f})~$\gamma=4.5$. Empty green squares are for $f=0.1$; filled purple circles are for $f=1$. The dashed lines represent the scaling with exponent $\widetilde{\gamma}=\gamma-1$ predicted by Equation~\eqref{eq:PK_scaling}; the dotted line is the scaling with exponent $\gamma$. Apart from preasymptotic effects for $\gamma$ close to $2$ and from finite-size effects for large values of $\gamma$, the scaling with exponent $\widetilde{\gamma}$ holds only up to\change{ the crossover value} $K^{*} \sim f\kmax$, while for $K>K^{*}$ the degree distributions follow the power-law of the backbones. The insets show that $\KMAX \sim f\kmax^{4-\gamma}$ for $2< \gamma \leq 3$ (panels (\textbf{a}--\textbf{c}) and ignoring logarithmic corrections for panel (\textbf{d})), and that $\KMAX \sim \theta \kmax$, for $\gamma>3$, where $\theta=1+f \langle r \rangle$ (panels (\textbf{e},\textbf{f})). Numerical simulations are performed by sampling $N$ nodes with degrees $k$ from $p(k)$, with $\kmin=3$ and $\kmax=\min\{N^{1/(\gamma-1)},N^{1/2}\}$, each of their $k$ neighbors excess degrees are sampled from $q(r)$, and the new connections created by the STC process are sampled from binomial distributions. The histograms of $P(K)$ are obtained for $N=10^8$; the insets are for various values of $N$. Results are averaged over 10 independent realizations.
		}
		\label{fig:degree_distribution}
	\end{figure}

	For power-law (PL) backbones with degree distribution $p(k) \sim k^{-\gamma}$, the generating functions have a singularity at $z=1$ given, at leading order in $\varepsilon=1-z$, by $g_0(1-\varepsilon) \sim 1 - c_1\varepsilon^{\gamma-1}$ and $ g_1(1-\varepsilon) \sim 1 - c_2\varepsilon^{\gamma-2}$,
	where $c_1, c_2$ are constants depending on $\gamma$ and in general on the low-degree part of the $p(k)$. With these expansions from Equation~\eqref{eq:G_0_STC} we get at leading order $G_0(1-\varepsilon) \sim 1 - C \varepsilon^{\gamma-2} = 1 - C
	\varepsilon^{\widetilde{\gamma}-1}$, where $\widetilde{\gamma}=\gamma-1$, and $C$ is a constant that depends also on $f$. Hence we can conclude, using asymptotic theorems for generating functions~\cite{flajolet1990singularity}, that\vspace{4pt}
	\begin{equation}
		P(K) \sim K^{-\widetilde{\gamma}}
		\label{eq:PK_scaling}
	\end{equation}
	for asymptotically large $K$. In other words, the STC mechanism decreases the exponent of the degree distribution by $1$. In Figure~\ref{fig:degree_distribution}, we report the results of numerical simulations for the degree distributions $P(K)$ of STC networks obtained from power-law backbones with $p(k) \sim k^{-\gamma}$, $\gamma>2$ and $\kmin \leq k \leq \kmax$, where $\kmax$ grows with the network size as $\kmax=\min\{ N^{1/(\gamma-1)},N^{1/2} \}$. Since we need to reach large sizes to check the theoretical asymptotic predictions, the simulations are performed by sampling nodes' neighborhoods instead of creating the entire STC graph.  Figure~\ref{fig:degree_distribution} clearly shows the scaling as $K^{-\widetilde{\gamma}}$, but, surprisingly, this scaling holds only up to $K^{*} \sim f \kmax$. For $K^{*}< K < \KMAX$ the exponent $\gamma$ is observed instead.

	\vspace{5pt}

	To solve this conundrum between theoretical predictions in the infinite-size limit and the double power-law scaling observed for finite-size networks, we identify two distinct mechanisms in the formation of large-degree nodes in $\mathcal{G}_f$. The first mechanism affects the degree distribution up to $K \sim f \kmax$, and it works as follows. Nodes of degree $\mathcal{O}(1)$ in $\mathcal{G}_0$ get a degree $K$ which is proportional to their neighbors excess degrees in $\mathcal{G}_0$. Since, in a finite system, these excess degrees are bounded by $\kmax$, a node of finite degree in $\mathcal{G}_0$ will have at most degree $K \sim f \kmax$ in $\mathcal{G}_f$. This explains the exponent $\widetilde{\gamma}=\gamma-1$ up to $K = K^{*} \sim f \kmax$. However, nodes in $\mathcal{G}_f$ can have a degree larger than $K^*$. Let us now consider hubs in $\mathcal{G}_0$, i.e., nodes with degree $k \sim \kmax$. Since $\kmax \gg 1$, we have, by the law of large numbers, $K \sim k + \sum_{i=1}^{k} fr_{i} \sim k (1+ f\langle r \rangle)$, which is a random variable following the original degree distribution $p(k)$, just rescaled by $\theta = 1+f\langle r \rangle$. Thus we have $P(K) \sim p(K/\theta)/\theta \sim K^{-\gamma}$, for $ \kmax < K < \KMAX$, where $\KMAX = \theta \kmax$. Note that $\theta=\mathcal{O}(1)$ for $\gamma>3$, implying that $\KMAX \sim \kmax$, while $\theta \sim \kmax^{3-\gamma}$ for $\gamma<3$, implying that $\KMAX \sim \kmax^{4-\gamma}$. Such a scaling of $\KMAX$ is confirmed in the insets of Figure~\ref{fig:degree_distribution} for various values of $\gamma$ and $f$.

	In conclusion, the theoretical prediction of the degree exponent $\widetilde{\gamma}$ holds in the infinite-size limit, but the degree distributions of finite-size STC networks exhibit a double power-law due to the fact that the degrees and the excess degrees in $\mathcal{G}_0$ are bounded by $\kmax$. This double power-law scaling of finite systems is also the reason why STC networks with backbone exponent $\gamma<3$---which are dense scale-free networks---do not violate graphicality constraints for scale-free degree distributions~\cite{delgenio2011all}. In fact, size-dependent cutoffs and crossovers soften the conditions for the graphicality of power-law degree sequences, \linebreak{}\mbox{see~\cite{baek2012fundamental,valigi2025graphicality}.}

	\section{Degree Correlations and Network Assortativity in STC Random~Graphs}
	\label{sec:correlations}
	
	A salient feature of real-world networks is the presence of degree correlations. Specifically, if $Q(K_1, K_2)$ is the probability that a uniformly randomly chosen link connects two nodes of degree $K_1$ and $K_2$, the presence of degree correlations means that $Q(K_1,K_2) \neq Q(K_1) Q(K_2)$.
	Here we discuss how degree correlations naturally arise in the STC model, and we provide a theory to quantify such correlations if the backbone graph itself is uncorrelated. Specifically we derive an exact expression for the Pearson correlation coefficient in infinite STC networks and show that its value is positive for any $f>0$, i.e., the STC mechanism always produces assortativity, regardless of the initial degree distribution $p(k)$. In the rest of this section we will often refer to the moments of this distribution, so for simplicity we denote these moments by $\mu_m = \langle k^m \rangle$.

	\subsection{The Pearson Correlation Coefficient}
	\label{corr:Pearson}
	
	A commonly used global measure of nearest-neighbor degree--degree correlations in networks is the Pearson coefficient $r$ (see Ref. \cite{newman2018networks}), which is the covariance of link end-node degrees normalized so that it falls in the range $r \in [-1, 1]$:
	\begin{align}
		\label{eq1.10}
		r = \frac{ \mathrm{Cov}(K_1, K_2) }{ \mathrm{Var}( \rightarrow K) },
	\end{align}
	%
	where the covariance is over all links in $\mathcal{G}_f$, in both directions, and $\mathrm{Var}( \rightarrow K)$ is the variance of link end-node degrees in $\mathcal{G}_f$, over all links in both directions. Here we derive $r$ in general and show that it is a function of only $f$ and the first four moments of the degree distribution $p(k)$ of $\mathcal{G}_0$.
	
		\subsubsection{Computation of $\mathrm{Var}( \rightarrow K)$}

	The variance in the denominator is straightforward to calculate, given that we know the generating function $G_0(z)$ of the degree distribution of $\mathcal{G}_f$, see Equation (\ref{eq:G_0_STC}). We can~write
	\begin{align}
		\label{eq1.20}
		\mathrm{Var}( \rightarrow K) &= \langle \rightarrow K^2 \rangle - \langle \rightarrow K \rangle^2  
		= \frac{ \langle K \rangle \langle K^3 \rangle - \langle K^2 \rangle^2 }{ \langle K \rangle^2 },
	\end{align}
	%
	where $\langle \rightarrow K \rangle$ and $\langle \rightarrow K^2 \rangle$ are the first two moments of the distribution of link end-node degrees in $\mathcal{G}_f$, and $\langle K \rangle$, $\langle K^2 \rangle$ and $\langle K^3 \rangle$ are the first three moments of the degree distribution of $\mathcal{G}_f$. We can express these three moments using the first three derivatives of $G_0(z)$ evaluated at $z=1$, and we find that they can be written in the form
	\begin{align}
		\langle K^n \rangle = \sum_{m=0}^n A_{n,m} f^m,  \label{eq1.30}
	\end{align}
	%
	where the coefficient $A_{n,m}$ is an expression (a Laurent polynomial) of only $\mu_1, \ldots, \mu_{n+1}$. Thus $\mathrm{Var}( \rightarrow K)$ is a function of $f, \mu_1, \mu_2, \mu_3, \mu_4$ and can be easily evaluated once we know the coefficients $A_{n,m}$ (see Appendix~\ref{sec:AP1} for details).

	\clearpage
	\subsubsection{Computation of $\mathrm{Cov}(K_1, K_2)$}
	
	{The covariance in the numerator of Equation (\ref{eq1.10}) may be most transparently expressed~as}
	\begin{align}
		\label{eq1.40}
		\mathrm{Cov}(K_1, K_2) = \langle K_1 K_2 \rangle_{Q(K_1, K_2)} - \langle \rightarrow K \rangle^2,
	\end{align}
	%
	where $\langle . \rangle_{Q(K_1, K_2)}$ is an average over the joint degree--degree distribution $Q(K_1,K_2)$ of $\mathcal{G}_f$. Although this joint distribution cannot be written explicitly, its generating function can be derived and the average in Equation (\ref{eq1.40}) evaluated. This derivation is rather involved, however; therefore, we do not present it here. Fortunately there is a considerably simpler approach to express the covariance in Equation (\ref{eq1.10}), by expanding it in terms of \mbox{conditional covariances.}
	
	The law of total covariance states that
	\begin{align}
		\mathrm{Cov}(X, Y) = \langle \mathrm{Cov}_{(X,Y)|Z} (X, Y) \rangle_Z + \mathrm{Cov}_Z ( \langle X \rangle_{(X,Y)|Z}, \langle Y \rangle_{(X,Y)|Z} ),  \nonumber
	\end{align}
	%
	where $\langle . \rangle_Z$ means averaging over the distribution of $Z$, $\langle . \rangle_{(X,Y) | Z}$ means averaging over the joint distribution of $(X,Y)$, given $Z$ and $\mathrm{Cov}_{(X,Y)|Z} (X, Y)$ is the covariance of $X$ and $Y$, over the joint distribution of $(X,Y)$, given $Z$. Let $\mathcal{T}$ denote the type of link: ``old'' or ``new''. The fraction of new links (see Ref. \cite{cirigliano2024strongly}) is given by
	\begin{align}
		\phi = \frac{ f \left( \langle k^2 \rangle - \langle k \rangle \right) }{ f \left( \langle k^2 \rangle - \langle k \rangle \right) + \langle k \rangle }.  \nonumber
	\end{align}
	%
	\indent Then the covariance we are looking for is
	\begingroup\makeatletter\def\f@size{8.9}\check@mathfonts
	\def\maketag@@@#1{\hbox{\m@th\fontsize{10}{10}\selectfont\normalfont#1}}
	\begin{align}
		\label{eq1.70}
		\mathrm{Cov}(K_1, K_2) &= \langle \mathrm{Cov}_{(K_1, K_2) | \mathcal{T}} (K_1, K_2) \rangle_{\mathcal{T}} + \mathrm{Cov}_{\mathcal{T}} ( \langle \rightarrow K_1 \rangle_{(K_1, K_2) | \mathcal{T}}, \langle \rightarrow K_2 \rangle_{(K_1, K_2) | \mathcal{T}} ) \nonumber \\
		&= \phi \, \mathrm{Cov}_{(K_1, K_2) | \mathrm{new}} (K_1, K_2) + (1-\phi) \, \mathrm{Cov}_{(K_1, K_2) | \mathrm{old}} (K_1, K_2) + \mathrm{Var}_{\mathcal{T}} ( \langle \rightarrow K \rangle_{(K_1,K_2)|\mathcal{T}} ),
	\end{align}
	\endgroup
	where $\langle \rightarrow K \rangle_{(K_1,K_2)|\mathcal{T}}$ is the mean end-node degree (over all links, both directions) of links of type $\mathcal{T}$, and $\mathrm{Var}_{\mathcal{T}} ( \langle \rightarrow K \rangle_{(K_1,K_2)|\mathcal{T}} )$ is the variance of that number over the two link types. All three terms in Equation (\ref{eq1.70}) can be explicitly written without direct knowledge of $Q(K_1,K_2)$ or its generating function. Here we briefly sketch the ideas behind their derivation, in particular the more complicated covariance terms, while the full exposition is presented in Appendix~\ref{sec:AP}.
	
	To compute the term $\mathrm{Cov}_{(K_1, K_2) | \mathrm{old}} (K_1, K_2)$ we write the new degree $K_1$ as the sum $K_1 = k_1 + \alpha_1 + \beta_1$, where $k_1$ is the original degree of the first end node, $\alpha_1$ is the number of new links that the first end node acquires via the other end node (connections to neighbors of the other end node) and $\beta_1$ is the number of new links that the first end node acquires via its neighbors that are not the other end node. We use analogous notation for the other end node: $K_2 = k_2 + \alpha_2 + \beta_2$. For a given backbone $\mathcal{G}_0$ the numbers $k_1$ and $k_2$ are constants while $\alpha_1, \alpha_2$ and $\beta_1, \beta_2$ are random variables. Importantly, due to the independence of the closure of each open triad in the STC process, these four random variables are mutually independent. To compute the term $\mathrm{Cov}_{(K_1, K_2) | \mathrm{old}} (K_1, K_2)$ we expand it using the law of total covariance again, now conditioning on the original degrees $k_1$ and $k_2$ of an old link. The result is a sum of two terms. The first is the mean of the covariance of $K_1$ and $K_2$ (averaged over the distribution of $k_1$ and $k_2$), which is zero due to the mutual independence of the random variables $\alpha_1, \alpha_2, \beta_1, \beta_2$. The second is the covariance of the means of $K_1$ and $K_2$ for given $k_1$ and $k_2$ which reduces to a variance due to the independence of $k_1$ and $k_2$; here we exploited the uncorrelated nature of the backbone $\mathcal{G}_0$.
	The end result is a strictly non-negative second-order polynomial in $f$, whose coefficients are expressions of $\mu_1, \mu_2, \mu_3$. See Appendix~\ref{sec:AP2} for details of the derivation. 	
	Computing the term $\mathrm{Cov}_{(K_1, K_2) | \mathrm{new}} (K_1, K_2)$ is a similar process. Now we write \linebreak{}$K_1 = k_1 + 1 + \beta_1 + \gamma_1$, where $\beta_1$ has the same meaning as before and the ``1'' term corresponds to the actual new link connecting the two end nodes. The random variable $\alpha_1$ is not present in this case, since the two end nodes of the new link were not neighbors initially, and the random variable $\gamma_1$ is the number of new neighbors the first end node acquires via the central node (the common neighbor of the two end nodes).
	Similarly to the old-link case we expand the covariance $\mathrm{Cov}_{(K_1, K_2) | \mathrm{new}} (K_1, K_2)$, conditioning on the original degrees $k_1$, $k_2$ of the two end nodes and $q$, the original degree of the central node. Once again, the first of the resulting terms, the mean of the covariance of $K_1$ and $K_2$ is zero due to the independence assumptions in the STC process. As in the old-link case the second term reduces to a variance for uncorrelated backbones.
	The end result is a strictly non-negative second-order polynomial in $f$, whose coefficients are expressions of $\mu_1, \mu_2, \mu_3, \mu_4$. See Appendix~\ref{sec:AP3} for details of the derivation.
	
	Finally, the ``type-mixing'' term $\mathrm{Var}_{\mathcal{T}} ( \langle \rightarrow K \rangle_{(K_1,K_2)|\mathcal{T}} )$ is an explicit variance, hence strictly non-negative. Computing it involves finding the first two moments of the distribution of final degree of a node of original degree $k$. See Appendix~\ref{sec:AP4} for details of \mbox{the derivation.}
	
	Substituting everything into Equation (\ref{eq1.10}), the Pearson coefficient can be written as a rational function,
	\begin{align}
		\label{eq1.72}
		r = \frac{ n_4 f^4 + n_3 f^3 + n_2 f^2 + n_1 f }{ d_4 f^4 + d_3 f^3 + d_2 f^2 + d_1 f + d_0 },
	\end{align}
	%
	where the $n_i$ and $d_i$ are polynomials of the moments $\mu_1$, $\mu_2$, $\mu_3$ and $\mu_4$ (see Appendix~\ref{sec:AP5} for the final expressions).

	\subsubsection{Positivity of the Pearson Coefficient}
	\label{corr:Pearson_pos}
	
	The Pearson coefficient is defined when the denominator in Equation (\ref{eq1.10}) is positive, which is always true apart from the pathological cases where $\mathcal{G}_0$ is degree-regular and $f=0$ or $f=1$.
	The sign of the Pearson coefficient is determined by the sign of the covariance in the numerator of Equation (\ref{eq1.10}). In the expansion of this covariance, Equation (\ref{eq1.70}), all three terms reduce to variances (see Appendix~\ref{sec:AP} for details), and are therefore strictly non-negative. The first two terms, the covariances for old and new links, are only zero when $\mathcal{G}_0$ is degree-regular. However, in this case, the third, ``type-mixing'' variance term is strictly positive, whenever $0<f<1$.
	
	To conclude, the STC mechanism produces strictly assortative degree--degree correlations for uncorrelated backbones, whenever the Pearson coefficient is defined.
	Qualitatively this may be understood as follows. Degree--degree correlations in the STC mechanism arise from the fact that both final end-node degrees of either an old or new link depend on the same old degrees (of end nodes of an old link or participant nodes of an old triad) in a positively correlated manner. Thus the mechanism that tends to produce a high degree for one end node of a link will also tend to produce a high degree for the other end node. Even in the case of a degree-regular original network, when end-node degrees are uncorrelated for old and new links separately, we still get assortative correlations overall, because the typical degree of nodes connected by old links will be different from that of new links [the third term in Equation (\ref{eq1.70})].

	\subsubsection{Behavior for Small $f$}
	\label{corr:small_f}

	When all four moments, $\mu_1, \mu_2, \mu_3, \mu_4$, are finite, we expand the Pearson coefficient in powers of $f$ and find that for small $f$ its behavior is given by
	\begin{align}
		\label{eq1.75}
		r &= \frac{n_1}{d_0} f + \mathcal{O}(f^2)  
		= \left[ 2 + \frac{ \mu_1 \mu_2 - \mu_1^2 }{ \mu_1 \mu_3 - \mu_2^2 }  \right] f + \mathcal{O}(f^2),
	\end{align}
	%
	where we assumed $d_0 \neq 0$, which is fulfilled for all but perfectly degree-regular networks.
	Since $\mu_2 \geq \mu_1$ and $\mu_1 \mu_3 \geq \mu_2^2$ for any degree distribution, the slope of $r$ at $f=0$ must be in the range $[2, \infty)$. Specifically, the slope diverges as the original network approaches a degree-regular network. The other extreme, a slope of $2$, is approached for heavy-tailed degree distributions. Below we deal with these two extreme cases in more detail separately.

	\subsubsection{Random Regular Network Backbones}
	\label{corr:RR}
	
\begin{figure}[H]
		\centering
		\includegraphics[width=0.97\textwidth]{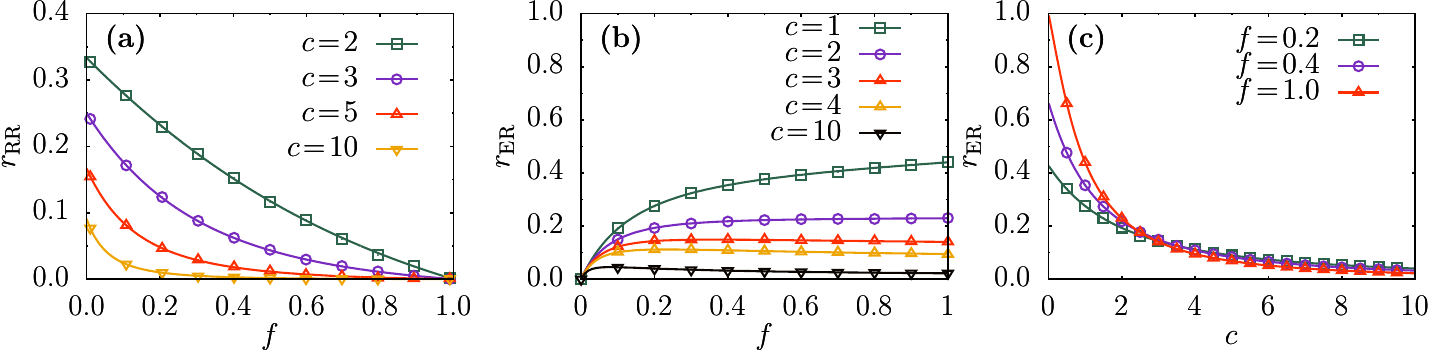}
		\caption{(\textbf{a}) Pearson correlation coefficient in STC networks generated from random regular networks, as a function of the STC parameter $f$, for different values of the degree parameter $c$. Panels (\textbf{b},\textbf{c}): Pearson correlation coefficient in STC networks generated from Erd\H{o}s--R\'enyi networks, (\textbf{b}) as a function of $f$, for different values of $c$, and (\textbf{c}) as a function of the mean degree $c$, for different values of the STC parameter $f$. Solid lines in all panels correspond to the exact results [Equation (\ref{eq1.77}) and Equation (\ref{eq1.80}) for panels (\textbf{a}) and (\textbf{b},\textbf{c}), respectively], and squares correspond to simulation results using networks of size $N=10^6$.}
		\label{fig:RR_ER_Pearson}
	\end{figure}

	Let us consider the special case of random regular networks, with all nodes in $\mathcal{G}_0$ having degree $c$. As noted earlier, Equation (\ref{eq1.75}) does not apply in this case. However, we now have $\mu_m = c^m$, so the full expression for the Pearson coefficient simplifies considerably,
	\begin{align}
		\label{eq1.77}
		\rRR(f,c) = \frac{ 1-f }{ (c^3 - 2c^2 + 1) f^2 + 2(c^2 - c - 1) f + (c+1) },
	\end{align}
	%
	which is confirmed in numerical simulations, see Figure~\ref{fig:RR_ER_Pearson}a.

	As previously noted, $r$ is not defined for $f=0$ in the case of random regular networks; however, the limit $f \to 0$ exists,
	\begin{align}
		\lim_{f \to 0} \rRR(f,c) = \frac{ 1 }{ c + 1 },  \nonumber
	\end{align}
	%
	which is verified by simulations using small $f$ values, see Figure~\ref{fig:RR_ER_Pearson}a.
	Interestingly, the Pearson coefficient for random regular networks decreases with $f$ near $f=0$, contrary to what happens for any other non-degenerate degree distribution $p(k)$ [see Equation (\ref{eq1.75}), and the example where $p(k)$ is a Poisson distribution, Figure~\ref{fig:RR_ER_Pearson}b].
	
	The case of $f=1$ is more subtle. The limiting behavior is
	\begin{align}
		\lim_{f \to 1} \rRR(f,c) = 0,  \nonumber
	\end{align}
	%
	which is verified by simulations in Figure~\ref{fig:RR_ER_Pearson}a. Exactly at $f=1$ the Pearson coefficient is undefined if we assume that every node in $\mathcal{G}_0$ has a perfect regular tree neighborhood up to second-nearest neighbors. This is not the case in random regular networks, however: in the infinite-size limit the expected number of short loops converges to a finite number. This means that a small disruption to the perfect regularity is expected in infinite random regular networks, meaning that the Pearson coefficient at $f=1$ is actually defined but has strong fluctuations. In this case $\rRR(f,c)$ is a random variable whose distribution converges to a limit distribution in the infinite-size limit, depending only on $c$, with a well-defined nonzero variance.

	\subsubsection{Erd\H{o}s--R\'enyi Network Backbones}
	\label{corr:ER}
	
	When the original network $\mathcal{G}_0$ is Erd\H{o}s--R\'enyi, i.e., has a Poisson degree distribution with mean degree $c$, the expression for the Pearson coefficient of $\mathcal{G}_f$ reduces to a relatively simple rational function,
	\begin{align}
		\label{eq1.80}
		\rER(f,c) = \frac{ c^2 f^4 + (2c^2+c) f^3 + 4c f^2 + 3f }{ c^4 f^4 + (5c^3+c^2) f^3 + (8c^2+2c) f^2 + (5c+2) f + 1 }.
	\end{align}
	%
	\indent {Figure~\ref{fig:RR_ER_Pearson}b,c confirms that this prediction coincides with simulation results. From Equation (\ref{eq1.80})} we see that the Pearson coefficient has well-defined limits as $c \to 0$ and $c \to \infty$,
	\begin{align}
		\lim_{c \to 0} \rER(f,c) = \frac{ 3f }{ 2f + 1 }, \quad \quad \lim_{c \to \infty} \rER(f,c) = 0,  \nonumber
	\end{align}
	%
	and that $\rER(f,c)$ is monotonically decreasing in $c$ for any $f$. Interestingly, we find that the behavior for small $f$, according to Equation (\ref{eq1.75}), is universal when the original network is Erd\H{o}s--R\'enyi, i.e.,
	\begin{align}
		\rER(f,c) = 3f + \mathcal{O}(f^2),  \nonumber
	\end{align}
	%
	independent of the mean degree.

	\subsubsection{Power-Law Backbone Networks}
	\label{corr:SF}

\begin{figure}[H]
		\centering
		\includegraphics[width=0.85\textwidth]{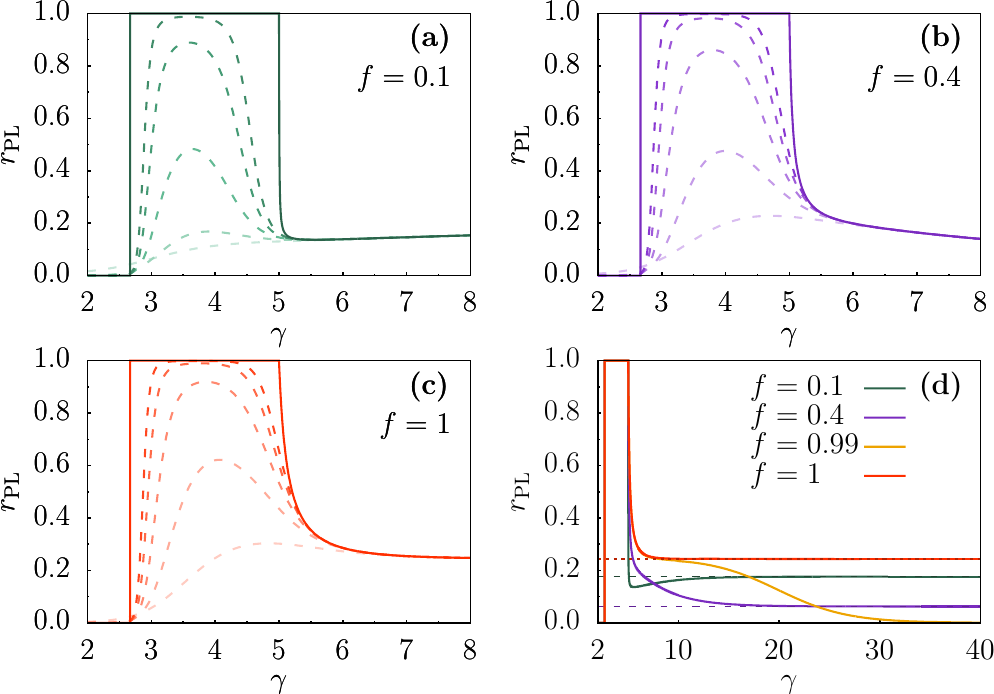}
		\caption{(\textbf{a}--\textbf{c}) Pearson correlation coefficient in STC networks generated from random power-law backbones, as a function of the degree distribution exponent $\gamma$, for various values of the STC parameter $f$. Dashed lines correspond to values of the Pearson coefficient evaluated numerically, using Equation (\ref{eq1.72}) and the $\kmax$-dependent expressions for the moments, for $\kmax=10^2, 10^3, 10^4, 10^5, 10^6$, darker colors for larger $\kmax$. The solid line corresponds to $k_{\textrm{max}} \to \infty$ in which case Equations (\ref{eq1.111}) and (\ref{eq1.113}) apply for $2 < \gamma < 5$, and for $\gamma > 5$, where the first four moments of the degree distribution are finite; these moments can be written and evaluated exactly, using the Hurwitz zeta function: $\langle k^n \rangle = \zeta(\gamma-n, k_{\textrm{min}}) / \zeta(\gamma, k_{\textrm{min}})$. (\textbf{d}) Pearson coefficients in the infinite-size limit, shown for large $\gamma$; dashed lines indicate the limits predicted by Equations (\ref{eq1.115}) and (\ref{eq1.116}). In all panels $k_{\textrm{min}}=3$.}
		\label{fig:PL_Pearson}
	\end{figure}

	When the network $\mathcal{G}_0$ has a power-law degree distribution, $p(k) = A k^{-\gamma}$ for $k_{\textrm{min}} \leq k \leq k_{\textrm{max}}$, we observe two sharp transitions in the Pearson coefficient, in the infinite-size limit (when $k_{\textrm{max}} \to \infty$), as a function of $\gamma$. Specifically, $r$ undergoes a discontinuous transition, jumping from $r=0$ to $r=1$ at $\gamma = 8/3$---with a nontrivial value $r \in (0,1)$ exactly at the threshold---and a continuous transition changing from $r=1$ to $r<1$ at $\gamma = 5$.
	The source of these transitions is that different moments of the degree distribution diverge, with $k_{\textrm{max}}$, in different ranges of $\gamma$. Specifically, the general formula for the Pearson coefficient involves the first four moments. For $\gamma > 5$ all these moments converge, and $r$ takes on a value in the open interval $(0,1)$. For $\gamma < 5$ some moments diverge, therefore one must carefully determine the dominant terms in the numerator and denominator of the expression in Equation (\ref{eq1.72}). For $3 \leq \gamma \leq 5$ the dominant terms in the numerator match the dominant terms in the denominator exactly, resulting in $r = 1$ in the limit $k_{\textrm{max}} \to \infty$. The range $(2,3)$ is more subtle: here all four moments diverge as powers of $k_{\textrm{max}}$ and the dominant terms change at an intermediate value, $\gamma = 8/3$, resulting in different asymptotics for $r$ when $\gamma < 8/3$ and $\gamma > 8/3$. A more detailed discussion of these results is presented in Appendix~\ref{sec:AP6}.
	The behavior of $r$ in the limit $k_{\textrm{max}} \to \infty$ can be summarized as follows,
	\vspace{-6pt}
	\begin{align}
		&2 < \gamma < 8/3: && r_{\textrm{PL}}(f, \gamma, k_{\textrm{min}})=0 \label{eq1.111} \\
		&\gamma = 8/3: && r_{\textrm{PL}}(f, \gamma, k_{\textrm{min}}) = \left( 1 + \frac{84}{ \zeta(5/3, k_{\text{min}}) }  \right)^{-1} \label{eq1.112} \\
		&8/3 < \gamma \leq 5: && r_{\textrm{PL}}(f, \gamma, k_{\textrm{min}})=1 \label{eq1.113} \\
		&5 < \gamma: && 0 < r_{\textrm{PL}}(f, \gamma, k_{\textrm{min}}) < 1 \label{eq1.114} \\
		& \gamma \to \infty: && r_{\textrm{PL}}(f<1, \gamma, k_{\textrm{min}})=\rRR(f<1, k_{\textrm{min}}) \label{eq1.115} \\
		& && r_{\textrm{PL}}(f=1, \gamma, k_{\textrm{min}})=\frac{3 k_{\textrm{min}} }{k_{\textrm{min}}^3 + k_{\textrm{min}}^2 + 1 },  \label{eq1.116}
	\end{align}
	%
	where $\zeta(\gamma, x) = \sum_{k=0}^{\infty} (k+x)^{-\gamma}$ is the Hurwitz zeta function.
	This behavior, presented in Figure~\ref{fig:PL_Pearson}a--c, is conceptually similar to that of the transitivity as a function of $\gamma$, in power-law STC networks \cite{cirigliano2024strongly}.

	The limit $\gamma \to \infty$ requires special attention. In this limit a random network of degree distribution $p(k) = A k^{-\gamma}$ converges to a random regular network of degree parameter $k_{\textrm{min}}$; therefore
	\begin{align}
		\lim_{\gamma \to \infty} r_{\textrm{PL}}(f, \gamma, k_{\textrm{min}}) = \rRR(f, k_{\textrm{min}}),  \nonumber
	\end{align}
	%
	for any $f<1$. This is true even in the limit $f \to 1$, i.e.,
	\begin{align}
		\lim_{f \to 1} \lim_{\gamma \to \infty} r_{\textrm{PL}}(f, \gamma, k_{\textrm{min}}) = \lim_{f \to 1} \rRR(f, k_{\textrm{min}}) = 0.  \nonumber
	\end{align}
	\indent However, setting $f=1$ first and then taking the limit $\gamma \to \infty$ results in a positive value, see Figure~\ref{fig:PL_Pearson}d. A simple approach to finding this limit is to substitute the degree distribution $p(k) = A k^{-\gamma}$ with a bimodal degree distribution on two possible degrees, $k_{\textrm{min}}$ and $k_{\textrm{min}}+1$. Specifically,
	\begin{align}
		p_{\mathrm{BM}}(k) =
		\begin{cases}
			k_{\textrm{min}} & \text{with probability } \rho = k_{\textrm{min}}^{-\gamma} / \left[ k_{\textrm{min}}^{-\gamma} + (k_{\textrm{min}}+1)^{-\gamma} \right] \\
			k_{\textrm{min}}+1 & \text{with probability } 1-\rho.
		\end{cases}
		\nonumber
	\end{align}
	\indent The distribution $p_{\mathrm{BM}}(k)$ converges to $p(k)$ for $\gamma \to \infty$, because the relative weight that $p(k)$ places on degrees $k > k_{\textrm{min}}+1$ vanishes as $\gamma \to \infty$. We can evaluate the first four moments of $p_{\mathrm{BM}}(k)$ and substitute into Equation~\eqref{eq1.72} with $f=1$. Then taking the limit $\rho \to 1$, which corresponds to the limit $\gamma \to \infty$, we get Equation~\eqref{eq1.116}.
	This discontinuous behavior of $r_{\textrm{PL}}$, with two different limits, is shown in Figure~\ref{fig:PL_Pearson}d where the Pearson coefficient is plotted for large values of $\gamma$. The curves for $f=0.1$, $f=0.4$ and $f=0.99$ approach the limits predicted by Equation~\eqref{eq1.115}. The curve for $f=1$ appears to almost coincide with the curve for $f=0.99$ up to $\gamma \approx 7$, but then converges to the value predicted by Equation (\ref{eq1.116}).

	\subsection{Average Nearest-Neighbor Degree}
	\label{corr:Knn}
	
\begin{figure}[H]
		\centering
		\includegraphics[width=0.85\textwidth]{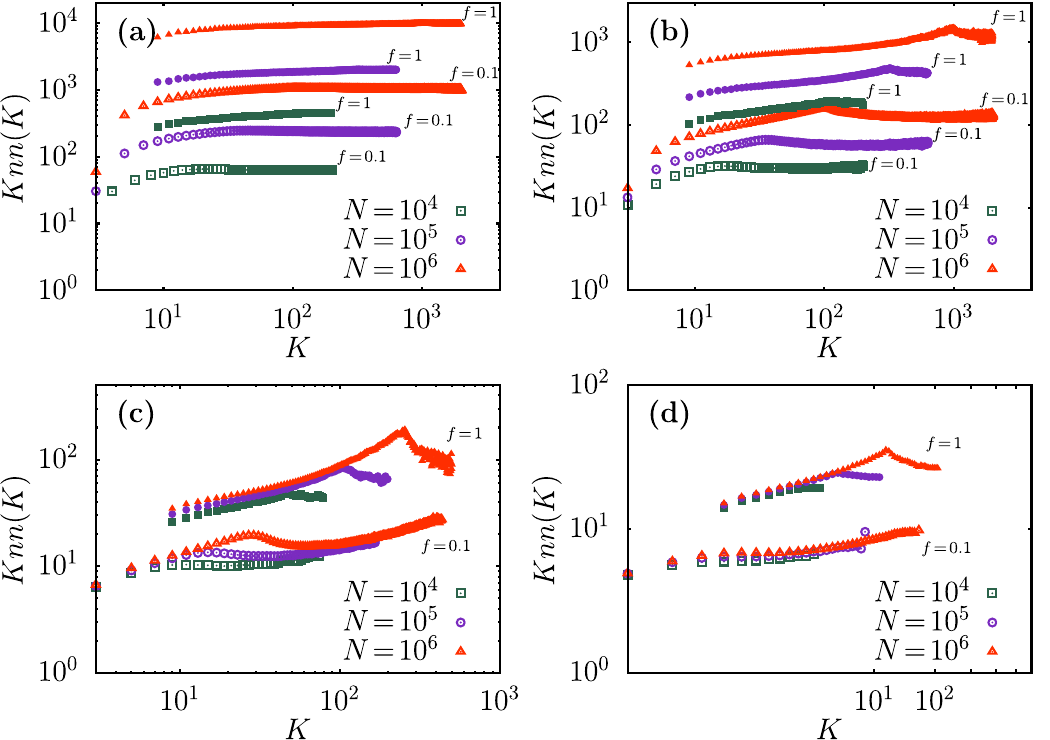}
		\caption{Average nearest-neighbor degree $K_{nn}(K)$ of STC random graphs with PL backbones with exponents (\textbf{a}) $\gamma=2.25$, (\textbf{b}) $\gamma=2.75$, (\textbf{c}) $\gamma=3.5$, (\textbf{d}) $\gamma=4.5$. The symbols represent numerical simulations on synthetic networks of size $N=10^4$ (squares), $N=10^5$ (circles) $N=10^6$ (triangles), with $\kmin=3$ and $\kmax=\min\{N^{1/(\gamma-1)}, N^{1/2} \}$. Empty symbols are for $f=0.1$; filled symbols are for $f=1$. The numerical simulations are performed using PL backbones created using the Uncorrelated Configuration Model~\cite{catanzaro2005generation}, creating the STC graphs by random closure of the triads. Results are averaged over 100 realizations for all data except $N=10^6$ in panel (\textbf{a}), where the average is over 10 realizations.
		}
		\label{fig:Knn_PL}
	\end{figure}

	To further characterize the mixing patterns in the STC networks, we analyze the average nearest-neighbor degree $K_{nn}(K)$, defined as the mean degree of neighbors of nodes with degree $K$~\cite{pastor2001dynamical}. This quantity can be computed from the conditional probability $P(K'|K)$ that a node of degree $K$ is connected to a node of degree $K'$
	\begin{equation}
		K_{nn}(K) = \sum_{K'} K' \, P(K'|K).
		\label{eq:Knn_def}
	\end{equation}
	%
	\indent Note that, by definition, we have 
	\begin{equation}
		\langle K_{nn}(K) \rangle_{Q(K)} = \langle K \rangle_{Q(K)} = \langle K^2 \rangle/ \langle K \rangle,
		\label{eq:Knn_normalization}
	\end{equation} 
	where $\langle \cdot \rangle$ denotes the averages with respect to the degree distribution $P(K)$, and $\langle \cdot \rangle_{Q(K)}$ denotes the averages with respect to $Q(K)=KP(K)/\langle K \rangle$. For uncorrelated networks, $P(K'|K)=Q(K)$ is independent of $K$, yielding a constant $K_{nn}(K) = \langle K^2 \rangle / \langle K \rangle$. Deviations from this behavior signal the presence of degree--degree correlations: an increasing $K_{nn}(K)$ with $K$ indicates assortative mixing, where high-degree nodes preferentially connect to other high-degree nodes, while a decreasing trend indicates disassortative behavior~\cite{newman2002assortative}. We can derive an elegant relation between the Pearson coefficient and $K_{nn}(K)$. From Equation~\eqref{eq1.10}, writing $Q(K',K)=Q(K)P(K'|K)$ and using Equation~\eqref{eq:Knn_def}, we have
	\begingroup\makeatletter\def\f@size{8.6}\check@mathfonts
	\def\maketag@@@#1{\hbox{\m@th\fontsize{10}{10}\selectfont\normalfont#1}}
	\begin{equation}
		r = \frac{\langle K' K \rangle_{Q(K',K)} - \langle K \rangle_{Q(K)} ^2}{ \langle K^2 \rangle_{Q(K)} - \langle K \rangle_{Q(K)}^2} = \frac{\sum_{K}KQ(K) \sum_{K'}K' P(K'|K) - \langle K \rangle_{Q(K)} ^2}{ \langle K^2 \rangle_{Q(K)} - \langle K \rangle_{Q(K)}^2} = \frac{\langle K_{nn}(K)K \rangle _{Q(K)} - \langle K \rangle_{Q(K)} ^2}{ \langle K^2 \rangle_{Q(K)} - \langle K \rangle_{Q(K)}^2}.
	\end{equation}
	\endgroup
	\indent Thus, the Pearson coefficient $r$ is related to a weighted average of $K_{nn}(K)$. This in turn implies that $K_{nn}(K)$ contains more information on the mixing patterns in a network than the Pearson coefficient: the price to pay for a larger amount of information is that we have no analytical expressions for $K_{nn}(K)$. However, we can compute $K_{nn}(K)$ numerically from simulations of synthetic STC networks. Results for PL backbones with various values of the exponent $\gamma$ and $f$ are presented in Figure~\ref{fig:Knn_PL}. Note the presence of a rapid change in the behavior of $K_{nn}(K)$ for $K > K^{*}$, mirroring the double power-law scaling observed in the degree distribution, see Figure~\ref{fig:degree_distribution}.

	For $\gamma=2.25$ in Figure~\ref{fig:Knn_PL}a, we see that $K_{nn}(K)$ depends weakly on $K$, in agreement with the vanishing Pearson coefficient in the infinite-size limit, Equation~\eqref{eq1.111}. However, we also see that $K_{nn}(K)$, at fixed $K$, grows with the system size, since the curves are shifted upward as $N$ increases. This effect can be explained using the normalization condition Equation~\eqref{eq:Knn_normalization}, together with an ansatz for $K_{nn}(K)$ of the form $K_{nn}(K) \sim a K^{\nu}$ for large $K$. Using this ansatz, we get 
	\begin{equation}
		K_{nn}(K) \sim \frac{\langle K^2 \rangle}{\langle K^{1+\nu} \rangle} K^{\nu},
		\label{eq:Knn_scaling_ansatz}
	\end{equation} 
	which grows with the system size whenever $\gamma<4$ we recall here that $\gamma$ is the degree exponent of the backbone, which implies that $\widetilde{\gamma}=\gamma-1$ is the degree exponent of $P(K)$, hence $\langle K^{n} \rangle$ diverges whenever $\gamma<2+n$ and $\nu<1$. Assuming $\nu \simeq 0$, consistent with $r=0$, we thus explain the presence of a multiplicative factor that grows with the \mbox{system size.} 
	
	For $\gamma=2.75$ in Figure~\ref{fig:Knn_PL}b, we see again that $K_{nn}(K)$ grows with the system size. However, we now have $r \to 1$ in the infinite-size limit since $2.75>8/3$, Equation~\eqref{eq1.113}. If we used Equation~\eqref{eq:Knn_scaling_ansatz} with $\nu=1$, we would expect a constant factor. One possible explanation for this behavior is the presence of enormous finite-size effects, which make it impossible to observe the true asymptotic behavior of $K_{nn}(K)$. Finite-size networks are weakly correlated, hence $K_{nn}(K)$ scales with an effective exponent $\nu \ll 1$, even though in the infinite-size limit $r \to 1$ and $\nu = 1$. This is in agreement with the behavior of the Pearson coefficient for finite $\kmax$, as shown in Figure~\ref{fig:PL_Pearson}, where a slow convergence to the infinite-$\kmax$ limit is observed.
	
	For $\gamma=3.5$ in Figure~\ref{fig:Knn_PL}c the size does not seem to play a strong role at fixed $K$, even though $\langle K^2 \rangle$ diverges. Indeed, using again Equation~\eqref{eq:Knn_scaling_ansatz} with $\nu \simeq 1$ we would expect a constant factor. The difference with the case $\gamma=2.75$ is that for $\gamma=3.5$ finite-size effects are much weaker: even finite-size networks are strongly correlated. In fact, as shown in Figure~\ref{fig:PL_Pearson}, $r$ approaches $1$ faster as $\kmax$ increases. 
	

	
	\section{The Clustering Spectrum in STC Random Graphs}
	\label{sec:clustering}
	
	The clustering is a measure of how likely nodes are to form triangles, hence it contains information on the loop structure of networks. Clustering in a network can be quantified in many ways. Two standard choices are the following. A global measure for the clustering is the ratio between the average number of triangles and the average number of triads, called transitivity, or global clustering coefficient. This is defined as~\cite{latora2017complex}
	\begin{equation}
		T = \frac{3 \langle N_{\triangle} \rangle}{\langle N_{\wedge} \rangle}.
	\end{equation}
	\indent The factor $3$ serves to normalize $T$, since each triangle contains three distinct triads. A local measure of the clustering of a given node $i$ can be defined as the number of triangles to which $i$ belongs divided by the number of triads centered on $i$, $C_i=n_{\triangle,i}/n_{\wedge,i}$~\cite{latora2017complex}. From this quantity, the average local clustering coefficient is obtained
	\begin{equation}
		\bar{C}=\frac{1}{N}\sum_{i=1}^{N} C_{i}.
	\end{equation}
	\indent Note that in general $T\neq\bar{C}$, as we can interpret $T$ as a weighted average of the local clustering coefficient $C_{i}$ with weights given by $n_{\wedge,i}$~\cite{latora2017complex}. Since a node of degree $K_i$ contributes $K_i(K_i-1)/2$ triads, we can condition on the degrees $K$ and define the clustering spectrum
	\begin{equation}
		\label{eq:clustering_spectrum}
		C(K)=\frac{2\mathbb{E}\left[ n_{\triangle}|K \right]}{K(K-1)},
	\end{equation}
	where $\mathbb{E}\left[ n_{\triangle}|K \right]$ is the average number of triangles to which a node of degree $K$ belongs. From the clustering spectrum we can compute $\bar{C}=\sum_{K}P(K)C(K)$, in the large network limit. Much more information about the local network structure is encoded in the clustering spectrum \cite{serrano2006clustering, yin2019local}. For this reason, it is in general much harder to find an expression for $C(K)$ in terms of the network properties.
	
	In~\cite{cirigliano2024strongly} we provided an exact expression for $T$ in STC random graphs with an arbitrary uncorrelated locally tree-like random backbone, and we compared $T$ with $\bar{C}$ measured in numerical simulations on synthetic STC networks. However, an analytical treatment of the full clustering spectrum $C(K)$ was missing. Here we fill this conceptual gap. By exploiting the local tree-likeness of the backbones $\mathcal{G}_0$, we provide exact formulae for the clustering spectra $C(K)$ in STC random graphs.
	
\begin{figure}[H]
		\centering
		\includegraphics[width=0.37\textwidth]{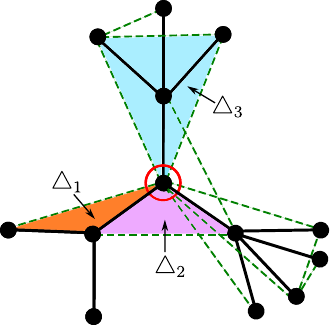}
		\caption{The three types of triangles formed by the Static Triadic Closure mechanism. The backbone $\mathcal{G}_0$ is represented with solid black lines; the connections created by the STC process are the green dashed lines. A node of original degree $k=3$ (red circle) forms new connections $\{\nu_1,\nu_2, \nu_3\}$ with its second neighbors such that its final degree is $K=k+\nu_1+\nu_2+\nu_3=9$. Such a node in $\mathcal{G}_f$ belongs to triangles of type 1 (orange), triangles of type 2 (pink), and triangles of type 3 (light blue). Given $k$ and $K$, the number of triangles of type 1 and 2 is fixed, while the number of triangles of type 3 depends only on the variables $\{ \nu_i\}$. The expected value of $n_{\triangle}$ conditioned on both $k$ and $K$ is then given by an average over the variables $\{\nu_i\}$ constrained to $\sum_{i=1}^{k}\nu_i = K-k$, see Equation~\eqref{eq:triangles_double_conditioned}.}
		\label{fig:triangles}
	\end{figure}

	Our task is to compute the conditional expectation $\mathbb{E}\left[ n_{\triangle}|K \right]$ in Equation~\eqref{eq:clustering_spectrum}. To do so, we can condition also on the value of the initial degree $k$
	\begin{equation}
		\label{eq:triangles_conditioned}
		\mathbb{E}\left[ n_{\triangle}|K \right] = \sum_{k}P(k|K)\mathbb{E}\left[ n_{\triangle}|k,K \right] = \frac{1}{P(K)}\sum_{k}p(k)P(K|k)\mathbb{E}\left[ n_{\triangle}|k,K \right].
	\end{equation}
	\indent Let us now work out explicitly the conditional expectation on the r.h.s. Given $k$ and $K$, let us consider all the possible realizations of the excess degrees $\{r_i \}_{i=1}^{k}$ and of the triadic closures $\{\nu_i \}_{i=1}^{k}$, where $\sum_{i=1}^{k}\nu_i = K-k$ is the number of new neighbors created by the STC process. Note that the $\nu_{i}$ are binomial variables distributed according to $\mathcal{B}(\nu_i|f,r_i)$, but we are constraining their sum to $K-k$. A node of original degree $k$ and final degree $K$ can be part of three types of triangles, depicted in Figure~\ref{fig:triangles}. There are $fk(k-1)/2$ triangles of type $\triangle_1$, $\sum_{i}\nu_i$ triangles of type $\triangle_2$, and $f\sum_{i}\nu_i(\nu_i-1)/2$ triangles of type $\triangle_3$. Thus we have
	\begin{align}
		\nonumber
		\mathbb{E}\left[n_{\triangle}|k,K\right] &= \sum_{\{r_i \}\{\nu_i \}}P(\{r_i \},\{\nu_i\}|k,K)\left[ f\frac{k(k-1)}{2} + \sum_{i=1}^{k}\nu_i \,+ f\sum_{i=1}^{k}\frac{\nu_i(\nu_i-1)}{2} \right]\\
		\label{eq:triangles_double_conditioned}
		&=\frac{fk(k-1)}{2}+K-k+\frac{f}{2}\sum_{\{\nu_i \}}P(\{\nu_i\}|k,K) \sum_{i=1}^{k}\nu_i(\nu_i-1),
	\end{align}
	where $\delta(x,y)$ is the Kronecker delta, and using the definition of conditional probability we can write
	\begin{equation}
		\label{eq:conditional_probability}
		P(\{\nu_i\}|k,K)  = \sum_{\{r_i\}}P(\{r_i \},\{\nu_i\}|k,K) = \frac{\sum_{\{r_i\}}P(\{r_i \},\{\nu_i\},K|k)}{P(K|k)},
	\end{equation}
	where, since the backbone is uncorrelated and locally tree-like,
	\begin{equation}
		\label{eq:full_probability}
		P(\{r_i\},\{\nu_i\},K|k) = \prod_{i=1}^{k}q(r_i)\mathcal{B}(\nu_i|r_i,f)\delta\left(K,k+\sum_{j=1}^{k}\nu_j\right),
	\end{equation}
	\indent In general, the last term on the r.h.s. of Equation~\eqref{eq:triangles_double_conditioned} can be evaluated via Monte Carlo sampling of the variables $k, \{r_i\}, \{\nu_i\}$, as well as for $P(K)$ and $P(K|k)$. Using Equation~\eqref{eq:triangles_conditioned} and \eqref{eq:clustering_spectrum} we then get the local clustering spectrum $C(K)$. Remarkably, as shown below, for some specific choices of the backbone degree distributions $p(k)$ it is possible to obtain some simpler expressions, and to extract the asymptotic scaling of $C(K)$ for large $K$.
	
	\subsection{Random Regular Backbones}
\begin{figure}[H]
		\centering
		\includegraphics[width=0.9\textwidth]{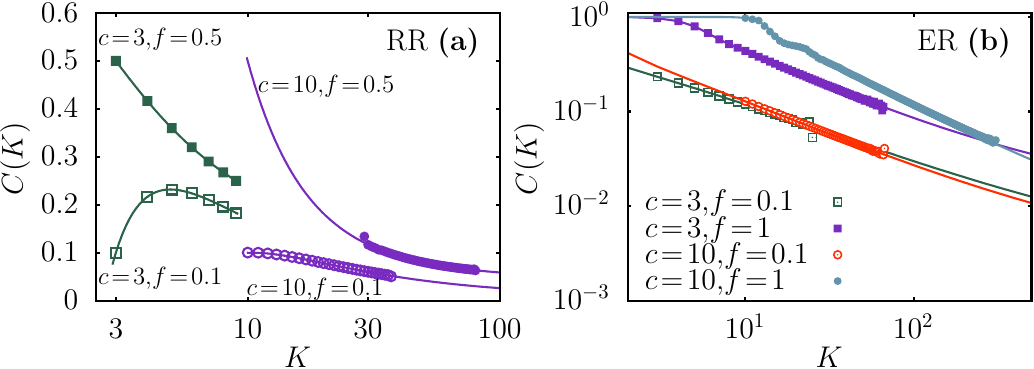}
		\caption{Clustering	spectra $C(K)$ of STC random graphs with RR backbones and ER backbones. (\textbf{a})~RR backbones with $c=3$ (squares) and $c=10$ (circles). Empty symbols are for $f=0.1$, filled symbols are for $f=0.5$. The lines are the theoretical prediction in Equation~\eqref{eq:clustering_RR}.
			(\textbf{b}) Results for ER backbones with $c=3$ (squares) and $c=10$ (circles). Empty symbols are for $f=0.1$, filled symbols are for $f=1$. The continuous lines are the numerical evaluation of Equation~\eqref{eq:clustering_ER}. 
			For both panels, STC networks of size $N=10^5$ have been generated by random STC process on the backbone, averaged over $100$ realizations. %
		}
		\label{fig:clustering_spectra_RR_ER}
	\end{figure}

	For RR backbones with connectivity $c$, whose degree distribution is $p(k)=\delta(k,c)$, explicit expressions can be obtained for the clustering spectrum $C(K)$.
	Let us consider the conditional distribution $P(\{n_i\}|k,K)$ as in Equation~\eqref{eq:conditional_probability}. The sum over the excess degrees $\{r_i\}$ is straightforward since $q(r)=\delta(r,c-1)$. The numerator gives
	\vspace{-6pt}
	\begin{align}
		\nonumber
		P(\{\nu_i\},K|k) &= \prod_{i=1}^{k}B(\nu_i|c-1,f)\delta\left(K,k+\sum_{j=1}^{k}\nu_j\right) \\
		&= \left[\prod_{i=1}^k {c-1 \choose \nu_i}\right] f^{K-k}(1-f)^{k(c-1)-(K-k)}\delta\left(K,k+\sum_{j=1}^{k}\nu_j\right),
	\end{align}
	while the denominator is simply given by
	\begin{equation}
		P(K|k)={k(c-1) \choose K-k} f^{K-k} (1-f)^{k(c-1)-(K-k)},
	\end{equation}
	because the variable $K-k$ is the sum of $k$ independent binomial random variables $\nu_i$, thus $S=K-k \sim \mathcal{B}(S|k(c-1),f)$. Putting these two pieces together we have
	\begin{equation}
		P(\{\nu_i\}|k,K) = \frac{\left[\prod_{i=1}^k {c-1 \choose \nu_i}\right] \delta\left(K,k+\sum_{j=1}^{k}\nu_j\right)}{{k(c-1) \choose K-k}},
	\end{equation}
	which is a multivariate hypergeometric distribution~\cite{feller1991introduction}. Over this distribution, we have
	\begin{equation}
		\mathbb{E}[\nu_i(\nu_i-1)] = \frac{(K-k)(K-k-1)(c-2)}{k[k(c-1)-1]},
	\end{equation}
	from which we can finally compute the clustering spectrum, using $P(k|K) = p(k)=\delta(k,c)$,
	\begin{equation}
		\label{eq:clustering_RR}
		C(K)=\frac{1}{K(K-1)}\left[fc(c-1)+2(K-c)+\frac{f(K-c)(K-c-1)(c-2)}{c(c-1)-1}\right].
	\end{equation}
	\indent We recall that, for RR backbones with finite $c$, the final degree $K$ is bounded by $c \leq K \leq c(c-1)$. Furthermore, for $f \to 1$ we have $P(K) \to \delta(K,c(c-1))$; thus $C(K) \to C(c(c-1))\delta(K,c(c-1))$. Results of numerical simulations on synthetic RR networks reported in Figure~\ref{fig:clustering_spectra_RR_ER}a show a perfect agreement with Equation~\eqref{eq:clustering_RR}. Interestingly, $C(K)$ can be monotone or non-monotone, depending on the value of $f$.
	
	\subsection{Erd\H{o}s--R\'enyi Backbones}
	The case of ER backbones, with a Poisson degree distribution $p(k)=e^{-c}c^k/k!$, is more complicated than the RR, yet analytical expressions for the clustering spectrum can be obtained. Let us focus on the conditional distribution $P(\{\nu_i\}|k,K)$ as in Equation~\eqref{eq:conditional_probability}. Let us start from the numerator
	\begin{equation}
		\sum_{\{r_i\}}P(\{r_i\},\{\nu_i\},K|k) = \sum_{\{r_i\}}\prod_{i=1}^{k}q(r_i)\mathcal{B}(\nu_i|r_i,f)\delta\left(K,k+\sum_{j=1}^{k}\nu_j\right),
	\end{equation}
	for the case of Poisson distribution $q(r)=e^{-c}c^{r}/r!$. The sum over the random variables $\{r_i \}$ gives, for each of the $i$-th terms,
	\begingroup\makeatletter\def\f@size{9.3}\check@mathfonts
	\def\maketag@@@#1{\hbox{\m@th\fontsize{10}{10}\selectfont\normalfont#1}}
	\begin{equation}
		\nonumber
		\sum_{r \geq \nu}q(r){r \choose \nu}f^{\nu}(1-f)^{r-\nu}=f^{\nu} \sum_{r \geq \nu} \frac{e^{-c}c^r}{r!}\frac{r!(1-f)^{r-\nu}}{\nu!(r-\nu)!} = \frac{(cf)^{\nu}}{\nu!}e^{-c}\sum_{r\geq \nu} \frac{[(1-f)c]^{r-\nu}}{(r-\nu)!} = \frac{(cf)^{\nu}}{\nu!}e^{-fc}.
	\end{equation}
	\endgroup
	\indent In other words, the convolution $\sum_{r}q(r)\mathcal{B}(\nu|r,f)$ when $q(r)$ is Poisson with average $c$, is a Poisson with average $fc$. Thus the variables $\nu_i$ are independent Poisson variables with average $fc$. The numerator then reads
	\begin{equation}
		\sum_{\{r_i\}}P(\{r_i\},\{\nu_i\},K|k) = \prod_{i=1}^{k}\frac{e^{-fc}(fc)^{\nu_i}}{\nu_i!}\delta\left(K,k+\sum_{i}\nu_i \right).
	\end{equation}
	\indent Coming back to the denominator in Equation~\eqref{eq:conditional_probability}, $P(K|k)$, the variable $K-k$, given $k$, is the sum of Poisson variables $K-k=\sum_{i=1}^{k}\nu_i$, thus it is Poisson with mean $kfc$. We can conclude that $P(K|k)$ follows a shifted Poisson distribution with mean $kfc$. Putting these two pieces together, we get
	\begin{equation}
		P(\{\nu_i\}|k,K) = \frac{\prod_{i=1}^{k}\frac{e^{-fc}(fc)^{\nu_i}}{\nu_i!}\delta \left(K,k+\sum_{i}\nu_i\right)}{\frac{e^{kfc}(kfc)^{K-k}}{(K-k)!}} = \frac{(K-k)!}{\prod_{i=1}^{k}\nu_i! k^{\nu_i}} \delta\left(K,k+\sum_{i}\nu_i\right),
	\end{equation}
	where we recognize a multinomial distribution~\cite{feller1991introduction} with all the probabilities being $1/k$ and the sum constrained to $K-k$. We can thus compute the expectation
	\begin{equation}
		\mathbb{E}\left[\sum_{i=1}^{k}n_i(n_i-1) \right] =  k \mathbb{E}\left[ n(n-1)\right] = k \frac{(K-k)(K-k-1)}{k^2} = \frac{(K-k)(K-k-1)}{k}.
	\end{equation}
	\indent Putting all these pieces together, we have
	\begingroup\makeatletter\def\f@size{9.8}\check@mathfonts
	\def\maketag@@@#1{\hbox{\m@th\fontsize{10}{10}\selectfont\normalfont#1}}
	\begin{equation}
		\label{eq:clustering_ER}
		C(K)=\frac{1}{K(K-1)}\sum_{k=1}^K \frac{p(k)p(K|k)}{\sum_{s=1}^K p(s)p(K|s)}\left[fk(k-1)+2(K-k)+f\frac{(K-k)(K-k-1)}{k} \right],
	\end{equation}
	\endgroup
	where $p(k)=e^{-c}c^k/k!$ and $p(K|k)=e^{-kfc} (kfc)^{(K-k)}/(K-k)!$. This average can be easily evaluated numerically. The results are shown in Figure~\ref{fig:clustering_spectra_RR_ER}b, in perfect agreement with numerical simulations on synthetic STC networks. Remarkably, an asymptotic expression for $C(K)$ for large $K$ can be obtained using a saddle-point evaluation of Equation~\eqref{eq:clustering_ER}. This procedure is described in Appendix~\ref{appendix:asymptotic}, and it gives
	\begin{equation}
		C(K) \sim \frac{f}{\left[A-2\log A \right]^2},
		\label{eq:CK_ER_asymptotics}
	\end{equation}
	where $A=1+fc+\log f + \log K$, for $K \gg 1$. Thus, the clustering spectrum decays to zero, albeit very slowly, and this decay would not be observable in simulations of networks of any feasible size. Interestingly, for moderate degree values $C(K)$ appears to decay approximately as a power-law, with exponent close to $-0.8$.

	\subsection{Power-Law Backbones}
\begin{figure}[H]
		\centering
		\includegraphics[width=0.85\textwidth]{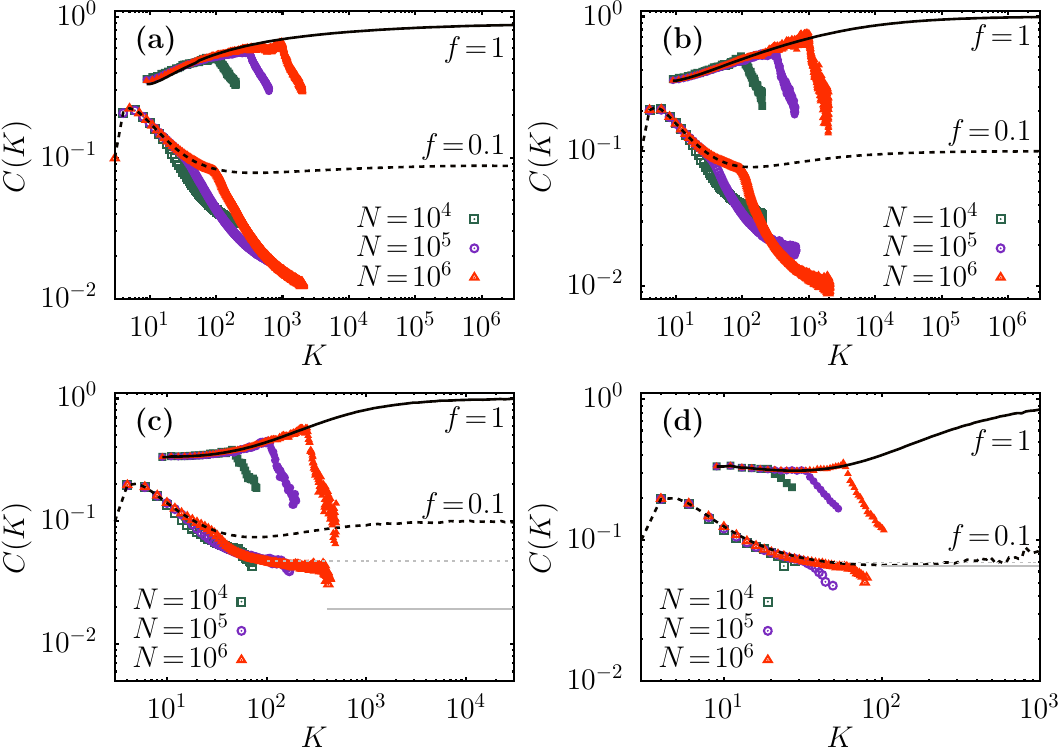}
		\caption{Clustering spectra $C(K)$ of STC random graphs with PL backbones with degree distribution exponents (\textbf{a}) $\gamma=2.25$, (\textbf{b}) $\gamma=2.75$, (\textbf{c}) $\gamma=3.5$, (\textbf{d}) $\gamma=4.5$. The symbols represent numerical simulations on synthetic networks of size $N=10^4$ (squares), $N=10^5$ (circles) $N=10^6$ (triangles), with $\kmin=3$ and $\kmax=\min\{N^{1/(\gamma-1)}, N^{1/2} \}$. Empty symbols are for $f=0.1$; filled symbols are for $f=1$. The black lines (dashed for $f=0.1$, continuous for $f=1$) are $C(K)$ as in Equation~\eqref{eq:clustering_spectrum} computed via Monte Carlo integration of Equation~\eqref{eq:triangles_double_conditioned}. Remarkably, these predict $C(K) \to f$ for $K \gg 1$ in infinite networks. The gray lines (dashed for $f=0.1$, continuous for $f=1$) in panels (\textbf{c},\textbf{d}) represent the value $f / ( 1 + f \langle r \rangle + f^2 \langle r \rangle^2 )$. The numerical simulations are performed using PL backbones created using the Uncorrelated Configuration Model~\cite{catanzaro2005generation}, creating the STC graphs by random closure of the triads. Results are averaged over 100 realizations for all data except $N=10^6$ in panel (\textbf{a}), where the average is over 10 realizations. The Monte Carlo integration is performed by random sampling of a node and its neighborhood, averaging over $10^8$ samples. The sharp decay of $C(K)$ observed in simulations for $K \sim K^* \sim f \kmax$ is due to finite-size effects, mirroring the behavior of the degree distributions, see Figure~\ref{fig:degree_distribution}.
		}
		\label{fig:clustering_spectra_PL}
	\end{figure}

	For STC graphs obtained from a PL backbone, we cannot derive an analytical expression for $C(K)$. However, we can numerically compute the sum in Equation~\eqref{eq:triangles_double_conditioned} by means of a Monte Carlo sampling, and compare it with numerical simulations on synthetic STC networks. The results are presented in Figure~\ref{fig:clustering_spectra_PL} for various values of $\gamma$ and $f$.
	The numerical simulations agree with the theoretical prediction up to degrees of approximately $K^* \sim f \kmax$, above which significant finite-size effects can be observed.
	Remarkably, in contrast to the case of homogeneous backbones---i.e., with exponentially bounded degree distributions---such as RR and ER, $C(K)$ does not decay to zero for $K \gg 1$. In fact, in infinite networks, it approaches a constant value $C(K) \sim f$ for $K\to \infty$. This implies that the hubs in $\mathcal{G}_f$, i.e., nodes with $K \sim K^{*} \gg 1$, are essentially surrounded by a clique---more precisely, a partial clique depending on the value of $f$. While we do not have a rigorous theoretical approach for deriving this scaling for $C(K)$ exactly, we can provide a heuristic argument to explain the behavior.

\begin{figure}[H]
		\centering
		\includegraphics[width=0.97\textwidth]{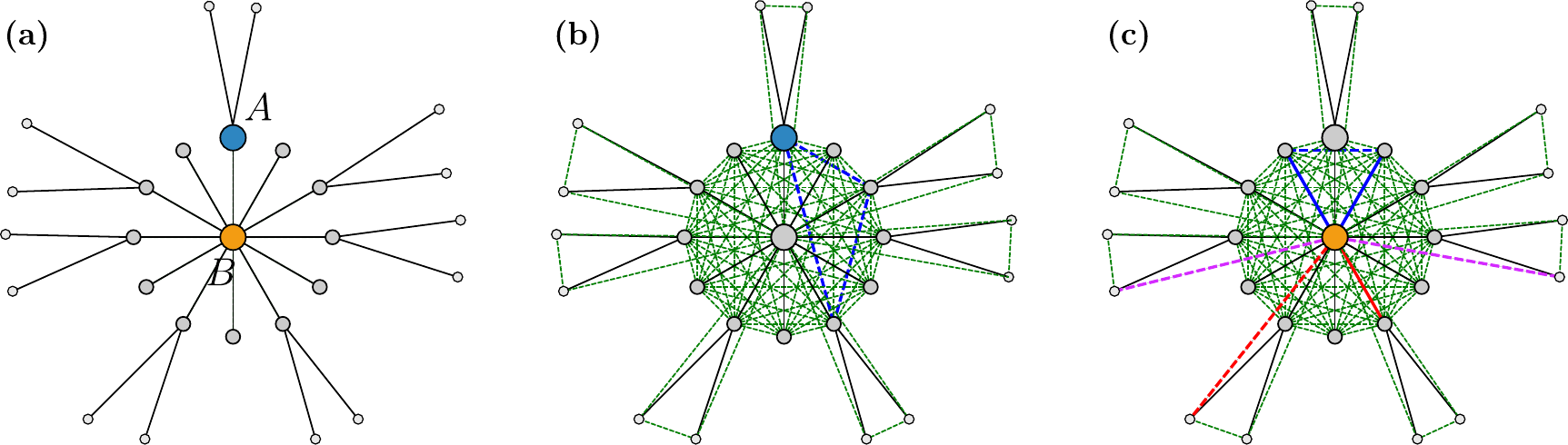}
		\caption{
			(\textbf{a}) A hub in $\mathcal{G}_0$ (node $B$--orange circle) with degree $\sim \kmax$ and its low-degree neighbor (node $A$--blue circle). (\textbf{b},\textbf{c}) The green dashed lines are the possible new edges that may be added by the STC process. (\textbf{b}) Out of all the possible types of triads centered on node $A$ in the network $\mathcal{G}_f$, the dominant type is a triad where the two end nodes are two of $A$'s original second neighbors via $B$ (highlighted by dashed blue lines). (\textbf{c}) Out of all the possible types of triads centered on node $B$ in the network $\mathcal{G}_f$, the three dominant types are the ones whose end nodes are (i) two original nearest neighbors of $B$ (highlighted by solid blue lines); (ii) one of $B$'s original nearest neighbors and one of $B$'s original second neighbors from a different branch (highlighted by solid/dashed red lines); and (iii) two original second neighbors of $B$ from different branches (highlighted by dashed purple lines).
		}
		\label{fig:CK_schematic}
	\end{figure}

	Consider a node $A$ with a small degree $k$ in $\mathcal{G}_0$ which is a nearest neighbor of a hub (node $B$, of degree $\sim \kmax$) in $\mathcal{G}_0$, see Figure~\ref{fig:CK_schematic}a. Node $A$ has a degree $K_A \sim k +f\kmax \sim K^*$ in $\mathcal{G}_f$. To estimate $C(K)$ for such a node, we need to consider the numbers of all the different types of triads, in $\mathcal{G}_f$, centered on node $A$, and the fraction of them that are closed. Without giving an exhaustive list of all the possible types of triads centered on node $A$, we observe that the dominant (most numerous) type is a triad whose end nodes are two of node $A$'s original second neighbors via node $B$ [dashed blue lines in Figure~\ref{fig:CK_schematic}b]. These triads dominate in number over all other types of triads centered on $A$ when $\kmax \to \infty$, and they are closed with probability $f$. This implies that $C(K) \to f$ for $K \sim K^*$.

	As seen in Figure~\ref{fig:clustering_spectra_PL}, the observed clustering spectra show considerable finite-size effects: $C(K)$ drops off at $K \sim K^*$ and decays to very low values for $\gamma < 3$ [Figure~\ref{fig:clustering_spectra_PL}a,b]. For $\gamma > 3$, after the drop-off, $C(K)$ appears to approach a nonzero value. We do not have a rigorous theoretical approach for deriving these effects, but we can repeat the above heuristic arguments to obtain a qualitative explanation.
	Let us consider a node $B$ which is already a hub in $\mathcal{G}_0$, with degree $\sim \kmax$, see Figure~\ref{fig:CK_schematic}a. Its final degree will be $K_B \sim \kmax(1+f\langle r \rangle) > K^{*}$.
	Once again, to estimate $C(K)$, we need to consider the numbers of all the different types of triads, in $\mathcal{G}_f$, centered on node $B$, and the fraction of them that are closed.
	Without listing all triad types, we observe three dominant types when $\kmax \to \infty$. The end nodes of these triads are (i) two original nearest neighbors of $B$ [blue lines in Figure~\ref{fig:CK_schematic}c]; (ii) one of $B$'s original nearest neighbors and one of $B$'s original second neighbors from a different branch [red solid/dashed lines in Figure~\ref{fig:CK_schematic}c]; and (iii) two original second neighbors of $B$ from different branches [purple dashed lines in Figure~\ref{fig:CK_schematic}c]. The respective numbers of these triads are (i) $\sim \kmax^2$, (ii) $\sim \kmax^2 \langle r \rangle f$, and (iii) $\sim \kmax^2 \langle r \rangle^2 f^2$. Type (i) triads are closed with probability $f$, while triads of type (ii) and (iii) are never closed. This implies that when $\gamma > 3$---so that $\langle r \rangle$ remains finite as $\kmax \to \infty$---we have approximately $C(K) \to f / ( 1 + f \langle r \rangle + f^2 \langle r \rangle^2 )$. Figure~\ref{fig:clustering_spectra_PL}c,d qualitatively supports the conclusion that $C(K)$ approaches a value between $0$ and $f$. When $\gamma < 3$, and $\langle r \rangle$ diverges with $\kmax$, we have $C(K) \to 0$, which is again supported by simulation results, see Figure~\ref{fig:clustering_spectra_PL}a,b.





	\section{Discussion}
	\label{sec:discussion}
	
	
	
	In this work we have examined the assortativity and local clustering properties of the Static Triadic Closure (STC) random network model~\cite{cirigliano2024strongly}, in which a clustered network is created via a triadic closure process on a backbone network. We consider the case of uncorrelated, locally tree-like backbones, in which triads are then closed independently with probability $f$ to give the final network.
	This allows us to obtain analytic expressions for the Pearson correlation coefficient and local clustering spectrum, exact in the infinite-size limit. These results were validated with extensive numerical simulations, which also allowed us to characterize the finite-size effects.

	The Pearson coefficient can be written as a function of the first four moments of the backbone degree distribution. Based on this general analysis, we give compact, exact expressions for random regular, Erd\H{o}s--R\'{e}nyi, and power-law degree distributed \mbox{backbone networks.}
	
	We found that the triadic closure process introduces assortative degree correlations; starting from an uncorrelated random network backbone, the final network (after triadic closure) always has positive correlations. Furthermore, this effect tends to increase with the fraction of triads closed, resulting in a direct positive correlation between degree assortativity and global clustering or transitivity.
	
	In the case of power-law degree distributions, the divergence of different moments of the backbone degree distribution leads to sharp transitions in the value of the Pearson coefficient. In particular, in the infinite-size limit, for any nonzero value of $f$, the Pearson coefficient (i) vanishes for degree distribution exponent $2 < \gamma < 8/3$, (ii) tends to $1$ for $8/3 < \gamma \leq 5$ and (iii) assumes a value $r \in [0, 1]$ for $\gamma > 5$. This behavior mirrors similar patterns seen in the transitivity in STC networks with power-law backbones~\cite{cirigliano2024strongly}.
	The mixing patterns in the network can also be characterized by the average nearest-neighbor degree (as a function of degree). We show that the Pearson coefficient can be written in terms of a weighted average of the average nearest-neighbor degree. Again, we find, through asymptotically exact expressions, nontrivial behavior, particularly with power-law degree-distributed backbones.
	
	We also find asymptotically exact formulae for the clustering spectrum, which contain summations over degree distributions. For the three cases of degree distributions studied these can easily be evaluated numerically.
	We show that the spectrum has nontrivial properties even for narrow backbone degree distributions.
	In the case of Erd\H{o}s--R\'{e}nyi backbone networks, the resulting clustering spectrum is strikingly power-law-like, decaying with an exponent close to $-0.8$, and is followed by a slower asymptotic decay for extremely large degrees.
	%
	When starting from a power-law degree-distributed backbone, the clustering spectrum is non-monotonic and exhibits strong finite-size effects.
	In the regime \mbox{$K \sim K^* \sim f \kmax$}, corresponding to low-degree backbone nodes adjacent to hubs, the dominant triads are closed with probability $f$, leading to $C(K) \to f$ as $\kmax \to \infty$. For degrees beyond $K^*$, associated with nodes that are already hubs in the backbone, $C(K)$ drops sharply; heuristic arguments suggest that it approaches a value between $0$ and $f$ for $\gamma > 3$, while it tends to zero for $\gamma \leq 3$.
	
	The analytical results we have obtained assume that the backbone network is a random, uncorrelated and locally tree-like network, defined by its degree distribution. However, most of the calculations may be adapted to weighted or degree--degree-correlated backbone networks, so long as the tree-like property is maintained. This would require adapting the equations to include weight probabilities for each edge, as well as, for the case of degree correlations, substituting sums over degrees weighted by the degree distribution with sums over edge degree--degree probabilities. Similarly, we believe that an analogous treatment should be possible for directed networks.
	
	Our results broadly agree with features frequently observed in real-world networks. Social networks, technological networks and certain biological networks such as protein--protein interaction networks are marked by significant transitivity (global clustering) and higher transitivity is associated with assortative degree--degree correlations
	\cite{newman2002assortative, newman2003social, foster2011clustering, filho2020transitivity}. Our results show that simple triadic closure is sufficient to produce this assortativity relation starting from uncorrelated backbone networks. Of course there are many other factors at play, leading to variability in assortativity, and in some cases negative correlations. Nevertheless, the positive correlations introduced naturally by triadic closure should be considered as a ``baseline''.
	Furthermore, real-world networks often have nontrivial clustering spectra. For instance, in networks with heavy-tailed degree distributions, clustering spectra exhibit slow decay with increasing degree \cite{stegehuis2017clustering, van2018triadic}. Again, the STC model naturally reproduces this effect.
	The analytical approach we have developed can be applied to STC networks resulting from  any locally tree-like backbone network, and the rich variety of results we have observed from the cases we have studied suggests that an appropriately chosen backbone structure may be able to reproduce more specific features of clustering spectra and assortativity observed in real-world networks of different types. We leave such a detailed study comparing real-world network features to corresponding STC model synthetic networks to future work.
	
	The STC model efficiently produces highly random synthetic networks containing complex patterns of overlapping short cycles, local clustering and degree assortativity. Remarkably, we are able to analytically calculate many local structural properties of STC networks. These networks exhibit varying patterns of such local properties, depending on the backbone network used. This suggests that more carefully tuned choices of backbone networks may be able to produce realistic patterns of a variety of global and local network structural properties, while retaining the tractability of STC networks.
	This underlines the value of the STC random network model in creating reference network ensembles, allowing analytical treatment of complex structures well beyond classical tree-like random \mbox{network models.}
	
	
	\vspace{6pt}
	
	
	
	
	

\acknowledgments{We thank Claudio Castellano for useful discussions in the early stage of this project. L.C. thanks the Department of Physics and i3N, University of Aveiro, for the warm hospitality. This work was developed partially within the scope of the project i3N, UIDB/50025/2020 \& UIDP/50025/2020, financed by national funds through the FCT/MEC. G.T. was supported by the Leverhulme Trust Project grant RPG-2023-187. L.C. was supported by PNRR MUR PE 0000023-NQSTI project grant. The raw data supporting the conclusions of this article will be made available by the authors on request.}



\appendix

\section[\appendixname~\thesection]{Computation of the Pearson Coefficient}
\label{sec:AP}

Here we present the final expressions, and their derivation, for the denominator in Equation~\eqref{eq1.10} and the three terms in the expansion of the covariance, in Equation (\ref{eq1.70}). We also present the final expression, in the form of a rational function, for the Pearson coefficient.

\subsection{Computation of $\mathrm{Var}( \rightarrow K)$}
\label{sec:AP1}

According to Equation (\ref{eq1.20}) we must compute the first three moments of the degree distribution of $\mathcal{G}_f$. The first three factorial moments are given by the first three derivatives of $G_0(z)$ evaluated at $z=1$, and thus we have
\begin{align}
	\langle K \rangle &= G_0'(1) \nonumber \\  
	\langle K^2 \rangle &= G_0''(1) + G_0'(1) \nonumber \\  
	\langle K^3 \rangle &= G_0'''(1) + 3 G_0''(1) + G_0'(1), \nonumber 
\end{align}
%
where $G_0(z) = g_0( z g_1(1-f+fz) )$.
The expression for $\langle K^n \rangle$, in general, can be written as
\begin{align}
	\langle K^n \rangle = \sum_{m=0}^n A_{n,m} f^m, \label{eqAP1.30}
\end{align}
%
where the coefficients for the first three orders are
\begin{align}
	A_{1,0} &= \mu_1  \nonumber \\  
	A_{1,1} &= \mu_2 - \mu_1  \nonumber \\  
	A_{2,0} &= \mu_2  \nonumber \\  
	A_{2,1} &= \frac{ 2 \mu_2^2 - \mu_1 \mu_2 - \mu_1^2 }{ \mu_1 }  \nonumber \\  
	A_{2,2} &= \frac{ \mu_1^3 + \mu_1^2 \mu_3 - 3 \mu_1 \mu_2^2 + \mu_2^3 }{ \mu_1^2 }  \nonumber \\  
	A_{3,0} &= \mu_3  \nonumber \\  
	A_{3,1} &= \frac{ -\mu_1^2 - 2 \mu_1 \mu_2 - 3 \mu_1 \mu_3 + 3 \mu_2^2 + 3 \mu_2 \mu_3 }{ \mu_1 }  \nonumber \\  
	A_{3,2} &= \frac{ 3 \mu_1^3 + 3 \mu_1^2 \mu_2 + 6 \mu_1^2 \mu_3 - 12 \mu_1 \mu_2^2 - 3 \mu_1 \mu_2 \mu_3 + 3 \mu_2^2 \mu_3}{ \mu_1^2 }  \nonumber \\  
	A_{3,3} &= \frac{ -2 \mu_1^4 - \mu_1^3 \mu_2 - 4 \mu_1^3 \mu_3 + \mu_1^3 \mu_4 + 9 \mu_1^2 \mu_2^2 - 3 \mu_1^2 \mu_2 \mu_3 + 2 \mu_1 \mu_2^3 - 3 \mu_2^4 + \mu_2^3 \mu_3 }{ \mu_1^3 },  \nonumber  
\end{align}
%
where $\mu_m = \langle k^m \rangle$ is the $m^{\textrm{th}}$ moment of the degree distribution of $\mathcal{G}_0$.
Using these coefficients, and Equation (\ref{eqAP1.30}), one can write $\mathrm{Var}( \rightarrow K)$ by substituting the moments $\langle K^n \rangle$ into Equation (\ref{eq1.20}).

\subsection{Computation of $\mathrm{Cov}_{(K_1, K_2) | \mathrm{old}} (K_1, K_2)$}
\label{sec:AP2}

\begin{figure}[H]
	\centering
	\includegraphics[width=0.45\textwidth]{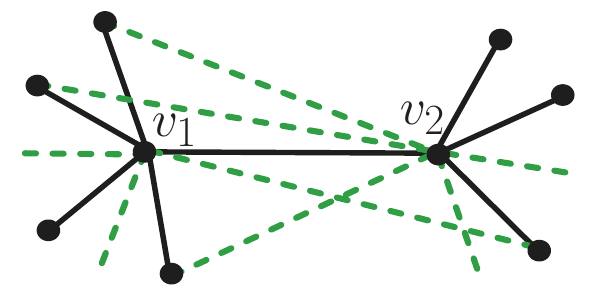}
	\caption{Schematic representation of the two classes of new links acquired by an end node of an old link. In this particular case we have $k_1=5, K_1=8, \alpha_1=1, \beta_1=2$ and $k_2=4, K_2=9, \alpha_2=3, \beta_2=2$.}
	\label{fig:cov_old}
\end{figure}

We expand the covariance for old links, conditioning on the original end-node degrees $k_1, k_2$:
\begin{align}
	\label{eqAP2.10}
	\mathrm{Cov}_{(K_1, K_2) | \mathrm{old}} (K_1, K_2) &= \langle \mathrm{Cov}_{(K_1, K_2) | (k_1,k_2)} (K_1, K_2) \rangle_{(k_1,k_2)} \nonumber \\
	& \quad \, + \mathrm{Cov}_{(k_1,k_2)} ( \langle \rightarrow K_1 \rangle_{(K_1, K_2) | (k_1,k_2)}, \langle \rightarrow K_2 \rangle_{(K_1, K_2) | (k_1,k_2)} ).
\end{align}
\indent We first show that $\mathrm{Cov}_{(K_1, K_2) | (k_1,k_2)} (K_1, K_2) = 0$ for any $(k_1,k_2)$, and thus its average over $(k_1,k_2)$---the first term in Equation (\ref{eqAP2.10})---is zero. To do this, let us write, for the final degree (in $\mathcal{G}_f$) of the two end nodes ($v_1$ and $v_2$) of an old link:
\begin{align}
	K_1 &= k_1 + \alpha_1 + \beta_1  \nonumber \\
	K_2 &= k_2 + \alpha_2 + \beta_2, \nonumber
\end{align}
%
where $k_1$ is a constant (original degree of node $v_1$), $\alpha_1$, a random variable, is the number of new links that node $v_1$ acquires via node $v_2$ (connections to neighbors of node $v_2$) and $\beta_1$, also a random variable, is the number of new links that node $v_1$ acquires via neighbors other than node $v_2$. The notation is analogous for $K_2$. (See Figure~\ref{fig:cov_old} for a schematic representation).

\vspace{-3pt}

The old degrees, $k_1$ and $k_2$, are constants, and, by the construction of the STC model, the four random variables, $\alpha_1, \alpha_2, \beta_1, \beta_2$, are mutually independent, so we have
\begin{align}
	\label{eqAP2.30}
	\mathrm{Cov}_{(K_1, K_2) | (k_1,k_2)} (K_1, K_2) = \mathrm{Cov}_{(K_1, K_2) | (k_1,k_2)} (k_1 + \alpha_1 + \beta_1, k_2 + \alpha_2 + \beta_2) = 0.
\end{align}
%
\indent Since this covariance is zero for any $k_1, k_2$, its average over the distribution of $(k_1, k_2)$ is also zero.
Now we pay attention to the second term in Equation (\ref{eqAP2.10}). Let us write the average final degree of the first end node of an old link, given the initial degrees $k_1, k_2$:
\begin{align}
	\langle \rightarrow K_1 \rangle_{(K_1, K_2) | (k_1,k_2)} &= k_1 + f(k_2-1) + f(k_1-1) \frac{ \langle k^2 \rangle - \langle k \rangle }{ \langle k \rangle } \nonumber \\
	&= \left( 1 + f \frac{ \langle k^2 \rangle - \langle k \rangle }{ \langle k \rangle } \right) k_1 + f k_2 - f \frac{ \langle k^2 \rangle }{ \langle k \rangle }. \nonumber
\end{align}
%
\indent Similarly,
\begin{align}
	\langle \rightarrow K_2 \rangle_{(K_1, K_2) | (k_1,k_2)} &= \left( 1 + f \frac{ \langle k^2 \rangle - \langle k \rangle }{ \langle k \rangle } \right) k_2 + f k_1 - f \frac{ \langle k^2 \rangle }{ \langle k \rangle },  \nonumber
\end{align}
%
i.e., both quantities are affine functions of $k_1$ and $k_2$. Using the bilinearity of covariance we can write the covariance of the above two quantities,
\begin{align}
	\label{eqAP2.60}
	&\mathrm{Cov}_{(k_1,k_2)} ( \langle \rightarrow K_1 \rangle_{(K_1, K_2) | (k_1,k_2)}, \langle \rightarrow K_2 \rangle_{(K_1, K_2) | (k_1,k_2)} ) = \nonumber \\
	& \quad =  f \left( 1 + f \frac{ \langle k^2 \rangle - \langle k \rangle }{ \langle k \rangle } \right) \left[ \mathrm{Cov}_{(k_1,k_2)}(k_1,k_1) + \mathrm{Cov}_{(k_1,k_2)}(k_2,k_2) \right] \nonumber \\
	& \quad \quad +  \left[ \left( 1 + f \frac{ \langle k^2 \rangle - \langle k \rangle }{ \langle k \rangle } \right)^2 + f^2 \right]   \mathrm{Cov}_{(k_1,k_2)}(k_1,k_2)  \\
	& \quad =  f \left( 1 + f \frac{ \langle k^2 \rangle - \langle k \rangle }{ \langle k \rangle } \right) \left[ \mathrm{Var}_{(k_1,k_2)}(k_1) + \mathrm{Var}_{(k_1,k_2)}(k_2) \right] \nonumber \\
	& \quad =  2 f \left( 1 + f \frac{ \langle k^2 \rangle - \langle k \rangle }{ \langle k \rangle } \right) \mathrm{Var}_{(k_1,k_2)}(k_1),\nonumber
\end{align}
%
where we exploited the symmetry and the factorization property (uncorrelated original network) of the joint distribution of $k_1$ and $k_2$, which is $p(k_1, k_2) = k_1 p(k_1) k_2 p(k_2) / \langle k \rangle^2 $. The variance of $k_1$, over the distribution of $(k_1,k_2)$, is
\vspace{-6pt}
\begin{align}
	\label{eqAP2.80}
	\mathrm{Var}_{(k_1,k_2)}(k_1) &= \langle k_1^2 \rangle_{(k_1,k_2)} - \langle k_1 \rangle_{(k_1,k_2)}^2  \nonumber \\
	&= \sum_{k_1=1}^{\infty} \sum_{k_2=1}^{\infty} \frac{ k_1^3 p(k_1) k_2 p(k_2) }{ \langle k \rangle^2 } - \left( \sum_{k_1=1}^{\infty} \sum_{k_2=1}^{\infty} \frac{ k_1^2 p(k_1) k_2 p(k_2) }{ \langle k \rangle^2 } \right)^2   \\
	&= \frac{ \langle k^3 \rangle \langle k \rangle -  \langle k^2 \rangle^2 }{ \langle k \rangle^2 }.\nonumber
\end{align}
%
\indent Using Equations (\ref{eqAP2.80}), (\ref{eqAP2.60}) and (\ref{eqAP2.30}) in Equation (\ref{eqAP2.10}) we finally we have the covariance we were seeking, written as a strictly non-negative second-order polynomial in $f$,
\begin{align}
	\label{eqAP2.100}
	\mathrm{Cov}_{(K_1, K_2) | \mathrm{old}} (K_1, K_2) &=  f \, \left[ 2 \, \frac{ \mu_3 \mu_1 - \mu_2^2 }{ \mu_1^2 } \right] + f^2 \, \left[ 2 \, \frac{ \mu_1 \mu_2 \mu_3 - \mu_3 \mu_1^2 - \mu_2^3 + \mu_2^2 \mu_1 }{ \mu_1^3 } \right] .
\end{align}
\subsection{Computation of $\mathrm{Cov}_{(K_1, K_2) | \mathrm{new}} (K_1, K_2)$}
\label{sec:AP3}

\begin{figure}[H]
	\centering
	\includegraphics[width=0.44\textwidth]{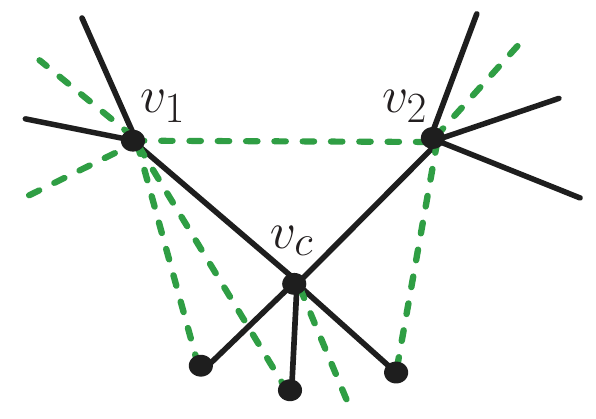}
	\caption{Schematic representation of the two classes of new links acquired by an end node of a new link. In this particular case we have $k_1=3, K_1=8, \beta_1=2, \gamma_1=2$ and $k_2=4, K_2=7, \beta_2=1, \gamma_2=1$.}
	\label{fig:cov_new}
\end{figure}

We will use the law of total covariance again here, conditioning on the original degrees $k_1, k_2$ of the end node of the given new link and the original degree $q$ of the central node of the triad that was closed:
\begin{align}
	\label{eqAP3.10}
	\mathrm{Cov}_{(K_1, K_2) | \mathrm{new}} (K_1, K_2) &= \langle \mathrm{Cov}_{(K_1, K_2) | (k_1,q,k_2)} (K_1, K_2) \rangle_{(k_1,q,k_2)} \nonumber \\
	& \quad \, + \mathrm{Cov}_{(k_1,q,k_2)} ( \langle \rightarrow K_1 \rangle_{(K_1, K_2) | (k_1,q,k_2)}, \langle \rightarrow K_2 \rangle_{(K_1, K_2) | (k_1,q,k_2)} ).
\end{align}
%
\indent We first show that $\mathrm{Cov}_{(K_1, K_2) | (k_1,q,k_2)} (K_1, K_2) = 0$ for any $(k_1,q,k_2)$, and thus its average over $(k_1,q,k_2)$---the first term in Equation (\ref{eqAP3.10})---is zero. To do this, similarly to Appendix~\ref{sec:AP2}, let us write, for the final degree (in $\mathcal{G}_f$) of the two end nodes ($v_1$ and $v_2$) of a new link:
\begin{align}
	K_1 &= k_1 + 1 + \beta_1 + \gamma_1  \nonumber \\
	K_2 &= k_2 + 1 + \beta_2 + \gamma_2, \nonumber
\end{align}
%
where $\beta_1$, a random variable, is the number of new links that node $v_1$ acquires via neighbors other than the central node, and $\gamma_1$ is a random variable indicating the number of new links via the central node (not counting the new link connecting to node $v_2$). The notation is analogous for $K_2$. (See Figure~\ref{fig:cov_new} for a schematic representation).

\vspace{-4pt}

The old degrees, $k_1$ and $k_2$, are constants, and, by the construction of the STC model, the four random variables, $\beta_1$, $\beta_2$, $\gamma_1$, $\gamma_2$, are mutually independent, so we have
\begingroup\makeatletter\def\f@size{9.5}\check@mathfonts
\def\maketag@@@#1{\hbox{\m@th\fontsize{10}{10}\selectfont\normalfont#1}}
\begin{align}
	\label{eqAP3.30}
	\mathrm{Cov}_{(K_1, K_2) | (k_1,q,k_2)} (K_1, K_2) = \mathrm{Cov}_{(K_1, K_2) | (k_1,q,k_2)} (k_1 + 1 + \beta_1 + \gamma_1, k_2 + 1 + \beta_2 + \gamma_2) = 0.
\end{align}
\endgroup
\indent Since this covariance is zero for any $k_1, q, k_2$, its average over the distribution of $(k_1, q, k_2)$ is also zero.
Now we deal with the second term in Equation (\ref{eqAP3.10}). We write the average final degree of the first end node of a new link, given the initial degrees $k_1,q, k_2$:
\begin{align}
	\langle \rightarrow K_1 \rangle_{(K_1, K_2) | (k_1,q,k_2)} &= k_1 + 1 + f(q-2) + f(k_1-1) \frac{ \langle k^2 \rangle - \langle k \rangle }{ \langle k \rangle } \nonumber \\
	&= \left( 1 + f \frac{ \langle k^2 \rangle - \langle k \rangle }{ \langle k \rangle } \right) k_1 + f q + 1 - f \frac{ \langle k^2 \rangle + \langle k \rangle }{ \langle k \rangle }.  \nonumber
\end{align}
%
\indent Similarly,
\begin{align}
	\langle \rightarrow K_2 \rangle_{(K_1, K_2) | (k_1,q,k_2)} &= \left( 1 + f \frac{ \langle k^2 \rangle - \langle k \rangle }{ \langle k \rangle } \right) k_2 + f q + 1 - f \frac{ \langle k^2 \rangle + \langle k \rangle }{ \langle k \rangle },  \nonumber
\end{align}
%
i.e., both quantities are affine functions of $k_1$, $k_2$ and $q$. Using the bilinearity of covariance we can write the covariance of the above two quantities,
\begin{align}
	\label{eqAP3.60}
	&\mathrm{Cov}_{(k_1,q,k_2)} ( \langle \rightarrow K_1 \rangle_{(K_1, K_2) | (k_1,q,k_2)}, \langle \rightarrow K_2 \rangle_{(K_1, K_2) | (k_1,q,k_2)} ) = \nonumber \\
	& \quad =   \left( 1 + f \frac{ \langle k^2 \rangle - \langle k \rangle }{ \langle k \rangle } \right)^2   \mathrm{Cov}_{(k_1,q,k_2)}(k_1,k_2) \nonumber \\
	& \quad \quad + f  \left( 1 + f \frac{ \langle k^2 \rangle - \langle k \rangle }{ \langle k \rangle } \right)  \left[ \mathrm{Cov}_{(k_1,q,k_2)}(k_1,q) + \mathrm{Cov}_{(k_1,q,k_2)}(k_2,q) \right]  \\
	& \quad \quad + f^2 \mathrm{Cov}_{(k_1,q,k_2)}(q,q) \nonumber \\
	& \quad =  f^2 \mathrm{Var}_{(k_1,q,k_2)}(q).\nonumber
\end{align}
%
where we used the fact that the joint distribution of $k_1, q, k_2$ factorizes:
\begin{align}
	p(k_1, q, k_2) = \frac{ q(q-1) p(q) k_1 p(k_1) k_2 p(k_2) }{ (\langle k^2 \rangle - \langle k \rangle) \langle k \rangle^2 }.  \nonumber
\end{align}
%
\indent The variance of $q$, over the distribution of $(k_1, q, k_2)$, is
\begingroup\makeatletter\def\f@size{8.7}\check@mathfonts
\def\maketag@@@#1{\hbox{\m@th\fontsize{10}{10}\selectfont\normalfont#1}}
\begin{align}
	\label{eqAP3.80}
	& \mathrm{Var}_{(k_1,q,k_2)}(q) = \langle q^2 \rangle_{(k_1,q,k_2)} - \langle q \rangle_{(k_1,q,k_2)}^2  \nonumber \\
	&= \sum_{k_1=1}^{\infty} \sum_{q=1}^{\infty} \sum_{k_2=1}^{\infty} \frac{ q^3(q-1) p(q) k_1 p(k_1) k_2 p(k_2) }{ (\langle k^2 \rangle - \langle k \rangle) \langle k \rangle^2 } - \left( \sum_{k_1=1}^{\infty} \sum_{q=1}^{\infty} \sum_{k_2=1}^{\infty} \frac{ q^2(q-1) p(q) k_1 p(k_1) k_2 p(k_2) }{ (\langle k^2 \rangle - \langle k \rangle) \langle k \rangle^2 } \right)^2   \\
	&= \frac{ \langle k^4 \rangle - \langle k^3 \rangle }{ \langle k^2 \rangle - \langle k \rangle } \, - \, \left( \frac{ \langle k^3 \rangle - \langle k^2 \rangle }{ \langle k^2 \rangle - \langle k \rangle } \right)^2.\nonumber
\end{align}
\endgroup
\indent Using Equations (\ref{eqAP3.80}), (\ref{eqAP3.60}) and (\ref{eqAP3.30}) in Equation (\ref{eqAP3.10}) we finally have the covariance we were seeking, written as a strictly non-negative second-order polynomial in $f$,
\begin{align}
	\mathrm{Cov}_{(K_1, K_2) | \mathrm{new}} (K_1, K_2) &= f^2 \, \frac{ \mu_4 \mu_2 + \mu_3 \mu_2 - \mu_4 \mu_1 + \mu_3 \mu_1 - \mu_3^2 - \mu_2^2 }{ \mu_2^2 - 2 \mu_2 \mu_1 + \mu_1^2 }.   \label{eqAP3.92}
\end{align}

\subsection{Computation of $\mathrm{Var}_{\mathcal{T}} ( \langle \rightarrow K \rangle_{(K_1,K_2)|\mathcal{T}} )$}
\label{sec:AP4}

Let us now deal with the variance term in Equation (\ref{eq1.70}), $\mathrm{Var}_{\mathcal{T}} ( \langle \rightarrow K \rangle_{(K_1,K_2)|\mathcal{T}} )$.
We must calculate the mean final degree of a node arrived at by following a random old link, and by following a random new link, respectively.
To help with calculations, we introduce $P(k,K)$, the probability that a uniformly randomly chosen node has old degree $k$ and final degree $K$. Then we have
\clearpage
\begin{align}
	P(k,K) = p(k) P(K|k).  \nonumber
\end{align}

\subsubsection{Mean final degree of node arrived at by following a random old link}

\noindent
\indent The probability that a node arrived at by following a random old link has old degree $k$ and final degree $K$ is
\begin{align}
	\frac{ k P(k,K) }{ \sum_{k=1}^{\infty} \sum_{K=1}^{\infty} k P(k,K) },  \nonumber
\end{align}

\noindent
so the probability that a node arrived at by following a random old link has final degree $K$~is
\begin{align}
	\frac{ \sum_{k=1}^{\infty} k P(k,K) }{ \sum_{k=1}^{\infty} \sum_{K=1}^{\infty} k P(k,K) }.  \nonumber
\end{align}

\noindent
\indent Therefore the mean final degree of a node arrived at by following a random old link is
\begin{align}
	\label{eqAP41.30}
	\langle \rightarrow K \rangle_{(K_1,K_2)|\textrm{old}} &= \frac{ \sum_{k=1}^{\infty} \sum_{K=1}^{\infty} K k P(k,K) }{ \sum_{k=1}^{\infty} \sum_{K=1}^{\infty} k P(k,K) }  \nonumber \\
	&= \frac{ \sum_{k=1}^{\infty} \sum_{K=1}^{\infty} K k p(k) P(K|k) }{ \sum_{k=1}^{\infty} \sum_{K=1}^{\infty} k p(k) P(K|k) }  \\
	&= \frac{ \sum_{k=1}^{\infty} k p(k) \sum_{K=1}^{\infty} K P(K|k) }{ \sum_{k=1}^{\infty} k p(k) \sum_{K=1}^{\infty} P(K|k) } \nonumber \\
	&= \frac{ \sum_{k=1}^{\infty} k p(k) \langle K \rangle_k }{ \langle k \rangle },\nonumber
\end{align}

\noindent
where $\langle K \rangle_k$ is the mean final degree of nodes of old degree $k$ and $\langle k \rangle$ is the mean degree of the original network $\mathcal{G}_0$. Fortunately, for the STC mechanism this can be easily written as
\begin{align}
	\label{eqAP41.40}
	\langle K \rangle_k = k (bf + 1),
\end{align}

\noindent
{where $b = ( \langle k^2 \rangle - \langle k \rangle ) / \langle k \rangle$ is the mean branching in $\mathcal{G}_0$. Using Equation (\ref{eqAP41.40}) in Equation (\ref{eqAP41.30})} we obtain
\begin{align}
	\label{eqAP41.70}
	\langle \rightarrow K \rangle_{(K_1,K_2)|\textrm{old}} = \frac{ \mu_2 }{ \mu_1 } + f \, \frac{ \mu_2^2 - \mu_2 \mu_1 }{ \mu_1^2 }.
\end{align}

\subsubsection{Mean final degree of node arrived at by following a random new link}

\noindent
\indent The probability that a node arrived at by following a random new link has old degree $k$ and final degree $K$ is
\begin{align}
	\frac{ (K-k) P(k,K) }{ \sum_{k=1}^{\infty} \sum_{K=1}^{\infty} (K-k) P(k,K) },  \nonumber
\end{align}

\noindent
so the probability that a node arrived at by following a random new link has final degree $K$~is
\begin{align}
	\frac{ \sum_{k=1}^{\infty} (K-k) P(k,K) }{ \sum_{k=1}^{\infty} \sum_{K=1}^{\infty} (K-k) P(k,K) }.  \nonumber
\end{align}

\noindent
\indent Therefore the mean final degree of a node arrived at by following a random new link is
\begin{align}
	\label{eqAP42.30}
	\langle \rightarrow K \rangle_{(K_1,K_2)|\textrm{new}} &= \frac{ \sum_{k=1}^{\infty} \sum_{K=1}^{\infty} K (K-k) P(k,K) }{ \sum_{k=1}^{\infty} \sum_{K=1}^{\infty} (K-k) P(k,K) } \nonumber \\
	&= \frac{  \sum_{k=1}^{\infty} \sum_{K=1}^{\infty} K^2 p(k) P(K|k)  -  \sum_{k=1}^{\infty} \sum_{K=1}^{\infty} Kk p(k) P(K|k)   }{ \sum_{k=1}^{\infty} \sum_{K=1}^{\infty} K p(k) P(K|k) - \sum_{k=1}^{\infty} \sum_{K=1}^{\infty} k p(k) P(K|k) }  \\
	&= \frac{  \sum_{k=1}^{\infty} p(k) \sum_{K=1}^{\infty} K^2 P(K|k)  -  \sum_{k=1}^{\infty} k p(k) \sum_{K=1}^{\infty} K P(K|k)   }{ \sum_{k=1}^{\infty} p(k) \sum_{K=1}^{\infty} K P(K|k) - \sum_{k=1}^{\infty} k p(k) \sum_{K=1}^{\infty} P(K|k) } \nonumber \\
	&= \frac{  \sum_{k=1}^{\infty} p(k) \langle K^2 \rangle_k  -  \sum_{k=1}^{\infty} k p(k) \langle K \rangle_k   }{ \sum_{k=1}^{\infty} p(k) \langle K \rangle_k - \langle k \rangle },\nonumber
\end{align}

\noindent
where $\langle K \rangle_k$ is the mean final degree of nodes of old degree $k$ [see Equation (\ref{eqAP41.40})] and $\langle K^2 \rangle_k$ is the second moment of the final degree distribution of nodes of old degree $k$.
To find $\langle K^2 \rangle_k$ let us write the final degree of a node of original degree $k$ as $K = k + \sum_{i=1}^k \nu_i$, where the $\nu_i$ are i.i.d. random variables whose distribution is generated by $g_1(1-f+fz)$. We can then write
\begin{align}
	K^2 = \left( k + \sum_{i=1}^k \nu_i \right)^2 = k^2 + 2 k \sum_{i=1}^k \nu_i + \sum_{i=1}^k \nu_i^2 + 2 \sum_{i=1}^{k-1} \sum_{j=i+1}^k \nu_i \nu_j,  \nonumber
\end{align}

\noindent
and so
\begin{align}
	\langle K^2 \rangle_k = k^2 + 2 k^2 \langle \nu \rangle + k \langle \nu^2 \rangle + (k^2 - k) \langle \nu \rangle^2.  \nonumber
\end{align}

\noindent
\indent We can use the generating function of the variables $\nu_i$ to find the first two moments,
\begin{align}
	\langle \nu \rangle &= f g_1'(1) = f \frac{ \langle k^2 \rangle - \langle k \rangle }{ \langle k \rangle }  \nonumber \\  
	\langle \nu^2 \rangle &= f^2 g_1''(1) + f g_1'(1) = f^2 \frac{ \langle k^3 \rangle - 3 \langle k^2 \rangle + 2 \langle k \rangle }{ \langle k \rangle } + f \frac{ \langle k^2 \rangle - \langle k \rangle }{ \langle k \rangle }, \nonumber  
\end{align}

\noindent
so we obtain
\begin{align}
	\label{eqAP42.90}
	\langle K^2 \rangle_k &= k^2 \left[ 1 + 2f \frac{ \langle k^2 \rangle - \langle k \rangle }{ \langle k \rangle } + f^2 \left( \frac{ \langle k^2 \rangle - \langle k \rangle }{ \langle k \rangle } \right)^2 \right]  \nonumber \\
	& \quad + k \left[ f^2 \frac{ \langle k^3 \rangle - 3 \langle k^2 \rangle + 2 \langle k \rangle }{ \langle k \rangle } + f \frac{ \langle k^2 \rangle - \langle k \rangle }{ \langle k \rangle } - f^2 \left( \frac{ \langle k^2 \rangle - \langle k \rangle }{ \langle k \rangle } \right)^2  \right]  ,
\end{align}

\noindent
and thus
\begin{align}
	\label{eqAP42.100}
	\sum_{k=1}^{\infty} p(k) \langle K^2 \rangle_k &= \langle k^2 \rangle \left[ 1 + 2f \frac{ \langle k^2 \rangle - \langle k \rangle }{ \langle k \rangle } + f^2 \left( \frac{ \langle k^2 \rangle - \langle k \rangle }{ \langle k \rangle } \right)^2 \right]  \nonumber \\
	& \quad + \langle k \rangle \left[ f^2 \frac{ \langle k^3 \rangle - 3 \langle k^2 \rangle + 2 \langle k \rangle }{ \langle k \rangle } + f \frac{ \langle k^2 \rangle - \langle k \rangle }{ \langle k \rangle } - f^2 \left( \frac{ \langle k^2 \rangle - \langle k \rangle }{ \langle k \rangle } \right)^2  \right]  ,
\end{align}

\noindent
\indent Using Equations (\ref{eqAP42.100}) and  (\ref{eqAP41.40}) in Equation (\ref{eqAP42.30}) we obtain
\begin{align}
	\label{eqAP42.130}
	\langle \rightarrow K \rangle_{(K_1,K_2)|\textrm{new}} = \frac{ \mu_2 + \mu_1 }{ \mu_1 } + f \, \frac{ \mu_3 \mu_1^2 + \mu_2^3 - 3 \mu_2^2 \mu_1 + \mu_1^3 }{ \mu_2 \mu_1^2 - \mu_1^3 }.
\end{align}

	\subsubsection{Constructing the full variance term}


Finally we can write the full variance term that appears in Equation (\ref{eq1.70}),
\begin{align}
	\label{eqAP43.10}
	\mathrm{Var}_{\mathcal{T}} ( \langle \rightarrow K \rangle_{(K_1,K_2)|\mathcal{T}} ) &= \phi \langle \rightarrow K \rangle_{(K_1,K_2)|\textrm{new}}^2 + (1-\phi) \langle \rightarrow K \rangle_{(K_1,K_2)|\textrm{old}}^2  \nonumber \\
	& \quad - \left[ \phi \langle \rightarrow K \rangle_{(K_1,K_2)|\textrm{new}} + (1-\phi) \langle \rightarrow K \rangle_{(K_1,K_2)|\textrm{old}} \right]^2   \\
	&= \phi (1-\phi) \left(  \langle \rightarrow K \rangle_{(K_1,K_2)|\textrm{new}} - \langle \rightarrow K \rangle_{(K_1,K_2)|\textrm{old}}   \right)^2.\nonumber
\end{align}

\subsection{Final Expression for the Pearson Coefficient}
\label{sec:AP5}

Collecting everything together, the Pearson coefficient, according to Equation (\ref{eq1.10}), can be written as a rational function
\begin{align}
	\label{eqAP5.10}
	r
	=
	\frac{
		n_4 f^4 + n_3 f^3 + n_2 f^2 + n_1 f
	}{
		d_4 f^4 + d_3 f^3 + d_2 f^2 + d_1 f + d_0,
	}
\end{align}
%
with the coefficients
\begin{align}
	n_4 &=
	\mu_1^6\mu_3 - \mu_1^6\mu_4 - \mu_1^5\mu_2^2
	+ 2\mu_1^5\mu_2\mu_4 - \mu_1^5\mu_3^2
	+ \mu_1^4\mu_2^3
	- \mu_1^4\mu_2^2\mu_3
	- \mu_1^4\mu_2^2\mu_4
	+ \mu_1^4\mu_2\mu_3^2,
	\nonumber \\[6pt]
	n_3 &=
	-\mu_1^7
	+ 2\mu_1^6\mu_2
	- \mu_1^6\mu_3
	+ \mu_1^6\mu_4
	- 5\mu_1^5\mu_2\mu_3
	- \mu_1^5\mu_2\mu_4
	+ 4\mu_1^4\mu_2^3
	+ 8\mu_1^4\mu_2^2\mu_3,
	\nonumber \\
	&\quad
	- 7\mu_1^3\mu_2^4
	- 2\mu_1^3\mu_2^3\mu_3
	+ 2\mu_1^2\mu_2^5,
	\nonumber \\[6pt]
	n_2 &=
	2\mu_1^7
	- 4\mu_1^6\mu_2
	- 2\mu_1^6\mu_3
	+ 4\mu_1^5\mu_2^2
	+ 6\mu_1^5\mu_2\mu_3
	- 6\mu_1^4\mu_2^3
	- 4\mu_1^4\mu_2^2\mu_3
	+ 4\mu_1^3\mu_2^4,
	\nonumber \\[6pt]
	n_1 &=
	-\mu_1^7
	+ 2\mu_1^6\mu_2
	+ 2\mu_1^6\mu_3
	- 3\mu_1^5\mu_2^2
	- 2\mu_1^5\mu_2\mu_3
	+ 2\mu_1^4\mu_2^3,
	\nonumber
\end{align}
for the numerator, and
\begin{align}
	d_4 &=
	\mu_1^7
	- 2\mu_1^6\mu_2
	+ 2\mu_1^6\mu_3
	- \mu_1^6\mu_4
	- 3\mu_1^5\mu_2^2
	- 3\mu_1^5\mu_2\mu_3
	+ 2\mu_1^5\mu_2\mu_4
	- \mu_1^5\mu_3^2,
	\nonumber \\
	&\quad
	+ 9\mu_1^4\mu_2^3
	+ 4\mu_1^4\mu_2^2\mu_3
	- \mu_1^4\mu_2^2\mu_4
	+ \mu_1^4\mu_2\mu_3^2
	- 9\mu_1^3\mu_2^4
	- 6\mu_1^3\mu_2^3\mu_3,
	\nonumber \\
	&\quad
	+ 7\mu_1^2\mu_2^5
	+ 4\mu_1^2\mu_2^4\mu_3
	- 4\mu_1\mu_2^6
	- \mu_1\mu_2^5\mu_3
	+ \mu_2^7,
	\nonumber \\[6pt]
	d_3 &=
	-3\mu_1^7
	+ 4\mu_1^6\mu_2
	- 8\mu_1^6\mu_3
	+ \mu_1^6\mu_4
	+ 13\mu_1^5\mu_2^2
	+ 16\mu_1^5\mu_2\mu_3
	- \mu_1^5\mu_2\mu_4,
	\nonumber \\
	&\quad
	- 28\mu_1^4\mu_2^3
	- 18\mu_1^4\mu_2^2\mu_3
	+ 25\mu_1^3\mu_2^4
	+ 14\mu_1^3\mu_2^3\mu_3
	- 15\mu_1^2\mu_2^5
	- 4\mu_1^2\mu_2^4\mu_3
	+ 4\mu_1\mu_2^6,
	\nonumber \\[6pt]
	d_2 &=
	3\mu_1^7
	- 3\mu_1^6\mu_2
	+ 9\mu_1^6\mu_3
	- 14\mu_1^5\mu_2^2
	- 20\mu_1^5\mu_2\mu_3
	+ 27\mu_1^4\mu_2^3
	+ 17\mu_1^4\mu_2^2\mu_3,
	\nonumber \\
	&\quad
	- 19\mu_1^3\mu_2^4
	- 6\mu_1^3\mu_2^3\mu_3
	+ 6\mu_1^2\mu_2^5,
	\nonumber \\[6pt]
	d_1 &=
	-\mu_1^7
	+ \mu_1^6\mu_2
	- 4\mu_1^6\mu_3
	+ 5\mu_1^5\mu_2^2
	+ 8\mu_1^5\mu_2\mu_3
	- 9\mu_1^4\mu_2^3
	- 4\mu_1^4\mu_2^2\mu_3
	+ 4\mu_1^3\mu_2^4,
	\nonumber \\[6pt]
	d_0 &=
	\mu_1^6\mu_3
	- \mu_1^5\mu_2^2
	- \mu_1^5\mu_2\mu_3
	+ \mu_1^4\mu_2^3, \nonumber
\end{align}
for the denominator.

\subsection{Pearson Coefficient for Power-Law Backbone Networks in the Infinite-Size Limit}
\label{sec:AP6}

We consider each region of $\gamma$ separately and determine the dominant terms in the expression for the Pearson coefficient, when $k_{\textrm{max}} \to \infty$.

\subsubsection{$5 < \gamma$}

In this range neither of the first four moments of the degree distribution diverge with $k_{\textrm{max}}$ and the Pearson coefficient assumes a value $0 < r < 1$.

	\subsubsection{$4 < \gamma \leq 5$}

In this range only $\mu_4$ diverges with $k_{\textrm{max}}$ and the dominant terms in the numerator match the dominant terms in the denominator exactly; therefore, we have $r=1$.

	\subsubsection{$3 < \gamma \leq 4$}

In this range both $\mu_3$ and $\mu_4$ diverge with $k_{\textrm{max}}$, but again the dominant terms in the numerator match the dominant terms in the denominator exactly; therefore, we still have $r=1$.

	\subsubsection{$\gamma = 3$}

In this case $\mu_2$ also diverges with $k_{\textrm{max}}$ (in addition to $\mu_3$ and $\mu_4$) but only logarithmically; therefore, it is sub-dominant and we still have $r=1$.

\subsubsection{$2 < \gamma < 3$}

In this range $\mu_2$, $\mu_3$ and $\mu_4$ all diverge as powers of $k_{\textrm{max}}$.
It will be informative to write the Pearson coefficient explicitly in this case.
To do this, we can use the continuous-degree approximation for the diverging moments,
\begin{align}
	\mu_m \sim \frac{ k_{\text{max}}^{m+1-\gamma} }{ (m+1-\gamma) \, \zeta(\gamma, k_{\text{min}}) },
	\label{eqAP5.20}
\end{align}

\noindent
and the exact formula for the convergent first moment,
\begin{align}
	\mu_1 = \frac{ \zeta(\gamma-1, k_{\text{min}}) }{ \zeta(\gamma, k_{\text{min}}) },
	\label{eqAP5.30}
\end{align}

\noindent
where $\zeta(\gamma, x) = \sum_{k=0}^{\infty} (k+x)^{-\gamma}$ is the Hurwitz zeta function.
The approximation in \mbox{Equation (\ref{eqAP5.20})} is asymptotically equivalent to the true value of $\mu_m$ for $k_{\text{max}} \to \infty$.
Using Equation (\ref{eqAP5.10}), keeping only dominant terms,
\begin{align}
	\label{eqAP5.40}
	r \sim \frac{ \mu_1^3 (\mu_4 \mu_2 - \mu_3^2) }{ \mu_1^3 (\mu_4 \mu_2 - \mu_3^2) + \mu_3 \mu_2^4 }.
\end{align}

\noindent
\indent Using Equations (\ref{eqAP5.20}) and (\ref{eqAP5.30}), this simplifies to
\begin{align}
	\label{eqAP5.50}
	r \sim \frac{ 1 }{ 1   +    \frac{  (5-\gamma)(4-\gamma)   }{  [ (3-\gamma)  \zeta(\gamma-1, k_{\text{min}})  ]^3   }      k_{\text{max}}^{8-3\gamma}      }.
\end{align}

\noindent
\indent As $k_{\text{max}} \to \infty$, $r$ tends to $0$ for $2 < \gamma < 8/3$, and $r \sim 1$ for $8/3 < \gamma \leq 3$. At the threshold value, $\gamma = 8/3$, we obtain a nontrivial solution,
\begin{align}
	\label{eqAP5.60}
	r \sim \frac{ 1 }{ 1   +    \frac{  84  }{  \zeta(5/3, k_{\text{min}})^3   }    }.
\end{align}


\section[\appendixname~\thesection]{{Asymptotic Scaling of the Clustering Spectrum for ER~Networks}}
\label{appendix:asymptotic}

\begin{figure}[H]
	\centering
	\includegraphics[width=0.44\textwidth]{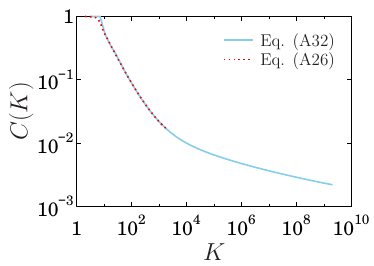}
	\caption{Comparison of $C(K)$, evaluated exactly according to Equation (\ref{eq20}), with the saddle-point approximation, Equation (\ref{eq:CK_ER_saddle_point}). The parameters used were $c=5$ and $f=1$.}
	\label{fig:CK_figure}
\end{figure}

We want to find the large-$K$ behavior
of the function $C(K)$ given in Equation~\eqref{eq:clustering_ER}. Let us write
\begin{align}
	\label{eq20}
	C(K) = \frac{1}{K(K-1)} \sum_{k=1}^K \frac{\Pi_K(k)}{\sum_{s=1}^K \Pi_K(s)} \left[ fk(k-1)+2(K-k)+f\frac{(K-k)(K-k-1)}{k} \right],
\end{align}
%
where, after some simplifications,
\begin{align}
	\label{eq30}
	\Pi_K(k) =\frac{ e^{-kfc} k^K }{ k! (K-k)! } (kf)^{-k}.
\end{align}
%
%
\indent For large $K$ the distribution $Q_K(k) = \Pi_K(k) / \sum_s \Pi_K(s)$ is sharply peaked around its average value, since it is a product of very quickly varying functions of $k$. Let us take the logarithm of Equation (\ref{eq30}). We have, using Stirling's formula for large $K$,
\begin{align}
	\nonumber
	\log \Pi_{K}(k) &= -kfc +K \log(k) -k \log(kf) - \log(k!)-\log[(K-k)!] \\
	&\sim -kfc +K\log(k) -k\log(kf) - k\log(k) -(K-k)\log(K-k) +K \\
	&=-K\Phi_K(k/K),\nonumber
\end{align}
where we introduced
\begin{equation}
	\Phi_K(x)\equiv -1+ x(fc+\log K +\log f ) - \log(x) + 2x \log x + (1-x) \log(1-x).
\end{equation}
\indent Finding the maximum $k^*(K)$ of $Q_K(k)$ is equivalent to finding the maximum $x^*(K)$ of $\Phi_K(x)$, via the relationship $k^*(K)=Kx^*(K)$.
Taking the derivative of $\Phi_K(x)$ we get
\begin{equation}
	\label{eq:saddle_point}
	\left.\frac{d \Phi(x)}{d x}\right|_{x*} = 0 \implies fc + \log K + \log f - \frac{1}{x^*} + 2 \log x^* +1 - \log(1-x^{*}) = 0.
\end{equation}
\indent Setting $y=1/x^*$ we get the transcendental equation
\begin{equation}
	y + \log(y) + \log(y-1) = 1 + fc + \log f + \log K \equiv A,
	\label{y_equation}
\end{equation}
which can be solved numerically. The solution of this equation gives us the saddle point $k^*(K)=K/y$, from which we can compute $C(K)$ as
\begin{equation}
	C(K) \sim \frac{1}{K(K-1)} \left[ f k^* (k^*-1) + 2(K-k^*) + f \frac{(K-k^*)(K-k^*-1)}{k^*} \right].
	\label{eq:CK_ER_saddle_point}
\end{equation}
\indent Figure \ref{fig:CK_figure} shows $C(K)$, for parameter values $c=5$ and $f=1$, evaluated exactly according to Equation (\ref{eq20}), compared with the saddle-point approximation, Equation (\ref{eq:CK_ER_saddle_point}), using the numerical solution of Equation (\ref{y_equation}). The exact evaluation is computationally demanding for large $K$; therefore, it was only done up to $K = 2 \times 10^3$. The perfect overlap of the curves, even for moderate values of $K$, confirms the validity of the saddle-point approximation.

\vspace{-4pt}

Since we are looking at $K \gg 1$, and $A \sim \log K \gg 1 $, the asymptotic behavior of $y$ is given by $y \sim A - 2 \log y$, from which it follows, neglecting terms of order $\log \log K / \log K$,~that
\begin{equation}
	y \sim A - 2 \log A.
	\label{eq:y}
\end{equation}
\indent Using Equations (\ref{eq:CK_ER_saddle_point}) and (\ref{eq:y}) we can write the leading term in $C(K)$ for large $K$ as
\begin{align}
	\label{eq1.15}
	C(K) &\sim \frac{ f k^* (k^*-1) }{ K(K-1) } \sim \frac{f}{y^2} \sim \frac{f}{[A-2\log A]^2},
\end{align}
where we recall that $A=1+fc+\log f + \log K$.

\bibliography{references}

\begin{thebibliography}{32}%
\makeatletter
\providecommand \@ifxundefined [1]{%
 \@ifx{#1\undefined}
}%
\providecommand \@ifnum [1]{%
 \ifnum #1\expandafter \@firstoftwo
 \else \expandafter \@secondoftwo
 \fi
}%
\providecommand \@ifx [1]{%
 \ifx #1\expandafter \@firstoftwo
 \else \expandafter \@secondoftwo
 \fi
}%
\providecommand \natexlab [1]{#1}%
\providecommand \enquote  [1]{``#1''}%
\providecommand \bibnamefont  [1]{#1}%
\providecommand \bibfnamefont [1]{#1}%
\providecommand \citenamefont [1]{#1}%
\providecommand \href@noop [0]{\@secondoftwo}%
\providecommand \href [0]{\begingroup \@sanitize@url \@href}%
\providecommand \@href[1]{\@@startlink{#1}\@@href}%
\providecommand \@@href[1]{\endgroup#1\@@endlink}%
\providecommand \@sanitize@url [0]{\catcode `\\12\catcode `\$12\catcode
  `\&12\catcode `\#12\catcode `\^12\catcode `\_12\catcode `\%12\relax}%
\providecommand \@@startlink[1]{}%
\providecommand \@@endlink[0]{}%
\providecommand \url  [0]{\begingroup\@sanitize@url \@url }%
\providecommand \@url [1]{\endgroup\@href {#1}{\urlprefix }}%
\providecommand \urlprefix  [0]{URL }%
\providecommand \Eprint [0]{\href }%
\providecommand \doibase [0]{https://doi.org/}%
\providecommand \selectlanguage [0]{\@gobble}%
\providecommand \bibinfo  [0]{\@secondoftwo}%
\providecommand \bibfield  [0]{\@secondoftwo}%
\providecommand \translation [1]{[#1]}%
\providecommand \BibitemOpen [0]{}%
\providecommand \bibitemStop [0]{}%
\providecommand \bibitemNoStop [0]{.\EOS\space}%
\providecommand \EOS [0]{\spacefactor3000\relax}%
\providecommand \BibitemShut  [1]{\csname bibitem#1\endcsname}%
\let\auto@bib@innerbib\@empty
\bibitem [{\citenamefont {Newman}(2018)}]{newman2018networks}%
  \BibitemOpen
  \bibfield  {author} {\bibinfo {author} {\bibfnamefont {M.~E.~J.}\
  \bibnamefont {Newman}},\ }\href
  {https://doi.org/10.1093/oso/9780198805090.001.0001} {\emph {\bibinfo {title}
  {{Networks}}}}\ (\bibinfo  {publisher} {Oxford University Press},\ \bibinfo
  {year} {2018})\BibitemShut {NoStop}%
\bibitem [{\citenamefont {Newman}(2002)}]{newman2002assortative}%
  \BibitemOpen
  \bibfield  {author} {\bibinfo {author} {\bibfnamefont {M.~E.~J.}\
  \bibnamefont {Newman}},\ }\bibfield  {title} {\bibinfo {title} {Assortative
  mixing in networks},\ }\href {https://doi.org/10.1103/PhysRevLett.89.208701}
  {\bibfield  {journal} {\bibinfo  {journal} {Phys. Rev. Lett.}\ }\textbf
  {\bibinfo {volume} {89}},\ \bibinfo {pages} {208701} (\bibinfo {year}
  {2002})}\BibitemShut {NoStop}%
\bibitem [{\citenamefont {Newman}\ and\ \citenamefont
  {Park}(2003)}]{newman2003social}%
  \BibitemOpen
  \bibfield  {author} {\bibinfo {author} {\bibfnamefont {M.~E.~J.}\
  \bibnamefont {Newman}}\ and\ \bibinfo {author} {\bibfnamefont
  {J.}~\bibnamefont {Park}},\ }\bibfield  {title} {\bibinfo {title} {Why social
  networks are different from other types of networks},\ }\href
  {https://doi.org/10.1103/PhysRevE.68.036122} {\bibfield  {journal} {\bibinfo
  {journal} {Phys. Rev. E}\ }\textbf {\bibinfo {volume} {68}},\ \bibinfo
  {pages} {036122} (\bibinfo {year} {2003})}\BibitemShut {NoStop}%
\bibitem [{\citenamefont {Foster}\ \emph {et~al.}(2011)\citenamefont {Foster},
  \citenamefont {Foster}, \citenamefont {Grassberger},\ and\ \citenamefont
  {Paczuski}}]{foster2011clustering}%
  \BibitemOpen
  \bibfield  {author} {\bibinfo {author} {\bibfnamefont {D.~V.}\ \bibnamefont
  {Foster}}, \bibinfo {author} {\bibfnamefont {J.~G.}\ \bibnamefont {Foster}},
  \bibinfo {author} {\bibfnamefont {P.}~\bibnamefont {Grassberger}},\ and\
  \bibinfo {author} {\bibfnamefont {M.}~\bibnamefont {Paczuski}},\ }\bibfield
  {title} {\bibinfo {title} {Clustering drives assortativity and community
  structure in ensembles of networks},\ }\href
  {https://doi.org/10.1103/PhysRevE.84.066117} {\bibfield  {journal} {\bibinfo
  {journal} {Phys. Rev E}\ }\textbf {\bibinfo {volume} {84}},\ \bibinfo {pages}
  {066117} (\bibinfo {year} {2011})}\BibitemShut {NoStop}%
\bibitem [{\citenamefont {Vasques~Filho}\ and\ \citenamefont
  {O'Neale}(2020)}]{filho2020transitivity}%
  \BibitemOpen
  \bibfield  {author} {\bibinfo {author} {\bibfnamefont {D.}~\bibnamefont
  {Vasques~Filho}}\ and\ \bibinfo {author} {\bibfnamefont {D.~R.~J.}\
  \bibnamefont {O'Neale}},\ }\bibfield  {title} {\bibinfo {title} {Transitivity
  and degree assortativity explained: The bipartite structure of social
  networks},\ }\href {https://doi.org/10.1103/PhysRevE.101.052305} {\bibfield
  {journal} {\bibinfo  {journal} {Phys. Rev. E}\ }\textbf {\bibinfo {volume}
  {101}},\ \bibinfo {pages} {052305} (\bibinfo {year} {2020})}\BibitemShut
  {NoStop}%
\bibitem [{\citenamefont {V{\'a}zquez}\ \emph {et~al.}(2002)\citenamefont
  {V{\'a}zquez}, \citenamefont {Pastor-Satorras},\ and\ \citenamefont
  {Vespignani}}]{vazquez2002large}%
  \BibitemOpen
  \bibfield  {author} {\bibinfo {author} {\bibfnamefont {A.}~\bibnamefont
  {V{\'a}zquez}}, \bibinfo {author} {\bibfnamefont {R.}~\bibnamefont
  {Pastor-Satorras}},\ and\ \bibinfo {author} {\bibfnamefont {A.}~\bibnamefont
  {Vespignani}},\ }\bibfield  {title} {\bibinfo {title} {Large-scale
  topological and dynamical properties of the internet},\ }\href
  {https://doi.org/10.1103/PhysRevE.65.066130} {\bibfield  {journal} {\bibinfo
  {journal} {Phys. Rev. E}\ }\textbf {\bibinfo {volume} {65}},\ \bibinfo
  {pages} {066130} (\bibinfo {year} {2002})}\BibitemShut {NoStop}%
\bibitem [{\citenamefont {Ravasz}\ and\ \citenamefont
  {Barab{\'a}si}(2003)}]{ravasz2003hierarchical}%
  \BibitemOpen
  \bibfield  {author} {\bibinfo {author} {\bibfnamefont {E.}~\bibnamefont
  {Ravasz}}\ and\ \bibinfo {author} {\bibfnamefont {A.-L.}\ \bibnamefont
  {Barab{\'a}si}},\ }\bibfield  {title} {\bibinfo {title} {Hierarchical
  organization in complex networks},\ }\href
  {https://doi.org/10.1103/PhysRevE.67.026112} {\bibfield  {journal} {\bibinfo
  {journal} {Phys. Rev. E}\ }\textbf {\bibinfo {volume} {67}},\ \bibinfo
  {pages} {026112} (\bibinfo {year} {2003})}\BibitemShut {NoStop}%
\bibitem [{\citenamefont {Serrano}\ and\ \citenamefont
  {Bogu\~n\'a}(2006)}]{serrano2006clustering}%
  \BibitemOpen
  \bibfield  {author} {\bibinfo {author} {\bibfnamefont {M.~A.}\ \bibnamefont
  {Serrano}}\ and\ \bibinfo {author} {\bibfnamefont {M.}~\bibnamefont
  {Bogu\~n\'a}},\ }\bibfield  {title} {\bibinfo {title} {Clustering in complex
  networks. i. general formalism},\ }\href
  {https://doi.org/10.1103/PhysRevE.74.056114} {\bibfield  {journal} {\bibinfo
  {journal} {Phys. Rev. E}\ }\textbf {\bibinfo {volume} {74}},\ \bibinfo
  {pages} {056114} (\bibinfo {year} {2006})}\BibitemShut {NoStop}%
\bibitem [{\citenamefont {Stegehuis}\ \emph {et~al.}(2017)\citenamefont
  {Stegehuis}, \citenamefont {van~der Hofstad}, \citenamefont {Janssen},\ and\
  \citenamefont {van Leeuwaarden}}]{stegehuis2017clustering}%
  \BibitemOpen
  \bibfield  {author} {\bibinfo {author} {\bibfnamefont {C.}~\bibnamefont
  {Stegehuis}}, \bibinfo {author} {\bibfnamefont {R.}~\bibnamefont {van~der
  Hofstad}}, \bibinfo {author} {\bibfnamefont {A.~J. E.~M.}\ \bibnamefont
  {Janssen}},\ and\ \bibinfo {author} {\bibfnamefont {J.~S.~H.}\ \bibnamefont
  {van Leeuwaarden}},\ }\bibfield  {title} {\bibinfo {title} {Clustering
  spectrum of scale-free networks},\ }\href
  {https://doi.org/10.1103/PhysRevE.96.042309} {\bibfield  {journal} {\bibinfo
  {journal} {Phys. Rev. E}\ }\textbf {\bibinfo {volume} {96}},\ \bibinfo
  {pages} {042309} (\bibinfo {year} {2017})}\BibitemShut {NoStop}%
\bibitem [{\citenamefont {van~der Hofstad}\ \emph {et~al.}(2018)\citenamefont
  {van~der Hofstad}, \citenamefont {van Leeuwaarden},\ and\ \citenamefont
  {Stegehuis}}]{van2018triadic}%
  \BibitemOpen
  \bibfield  {author} {\bibinfo {author} {\bibfnamefont {R.}~\bibnamefont
  {van~der Hofstad}}, \bibinfo {author} {\bibfnamefont {J.~S.~H.}\ \bibnamefont
  {van Leeuwaarden}},\ and\ \bibinfo {author} {\bibfnamefont {C.}~\bibnamefont
  {Stegehuis}},\ }\bibfield  {title} {\bibinfo {title} {Triadic closure in
  configuration models with unbounded degree fluctuations},\ }\href
  {https://doi.org/10.1007/s10955-018-1952-x} {\bibfield  {journal} {\bibinfo
  {journal} {J. Stat. Phys.}\ }\textbf {\bibinfo {volume} {173}},\ \bibinfo
  {pages} {746} (\bibinfo {year} {2018})}\BibitemShut {NoStop}%
\bibitem [{\citenamefont {Li}\ \emph {et~al.}(2017)\citenamefont {Li},
  \citenamefont {Shang},\ and\ \citenamefont {Yang}}]{li2017clustering}%
  \BibitemOpen
  \bibfield  {author} {\bibinfo {author} {\bibfnamefont {Y.}~\bibnamefont
  {Li}}, \bibinfo {author} {\bibfnamefont {Y.}~\bibnamefont {Shang}},\ and\
  \bibinfo {author} {\bibfnamefont {Y.}~\bibnamefont {Yang}},\ }\bibfield
  {title} {\bibinfo {title} {Clustering coefficients of large networks},\
  }\href {https://doi.org/10.1016/j.ins.2016.12.027} {\bibfield  {journal}
  {\bibinfo  {journal} {Inf. Sci.}\ }\textbf {\bibinfo {volume} {382}},\
  \bibinfo {pages} {350} (\bibinfo {year} {2017})}\BibitemShut {NoStop}%
\bibitem [{\citenamefont {Van Der~Hofstad}\ \emph {et~al.}(2017)\citenamefont
  {Van Der~Hofstad}, \citenamefont {Janssen}, \citenamefont {Van~Leeuwaarden},\
  and\ \citenamefont {Stegehuis}}]{van2017local}%
  \BibitemOpen
  \bibfield  {author} {\bibinfo {author} {\bibfnamefont {R.}~\bibnamefont {Van
  Der~Hofstad}}, \bibinfo {author} {\bibfnamefont {A.~J. E.~M.}\ \bibnamefont
  {Janssen}}, \bibinfo {author} {\bibfnamefont {J.~S.~H.}\ \bibnamefont
  {Van~Leeuwaarden}},\ and\ \bibinfo {author} {\bibfnamefont {C.}~\bibnamefont
  {Stegehuis}},\ }\bibfield  {title} {\bibinfo {title} {Local clustering in
  scale-free networks with hidden variables},\ }\href
  {https://doi.org/10.1103/PhysRevE.95.022307} {\bibfield  {journal} {\bibinfo
  {journal} {Phys. Rev. E}\ }\textbf {\bibinfo {volume} {95}},\ \bibinfo
  {pages} {022307} (\bibinfo {year} {2017})}\BibitemShut {NoStop}%
\bibitem [{\citenamefont {Holme}\ and\ \citenamefont
  {Kim}(2002)}]{holme2002growing}%
  \BibitemOpen
  \bibfield  {author} {\bibinfo {author} {\bibfnamefont {P.}~\bibnamefont
  {Holme}}\ and\ \bibinfo {author} {\bibfnamefont {B.~J.}\ \bibnamefont
  {Kim}},\ }\bibfield  {title} {\bibinfo {title} {Growing scale-free networks
  with tunable clustering},\ }\href
  {https://doi.org/10.1103/PhysRevE.65.026107} {\bibfield  {journal} {\bibinfo
  {journal} {Phys. Rev. E}\ }\textbf {\bibinfo {volume} {65}},\ \bibinfo
  {pages} {026107} (\bibinfo {year} {2002})}\BibitemShut {NoStop}%
\bibitem [{\citenamefont {Serrano}\ and\ \citenamefont
  {Bogu\~n\'a}(2005)}]{angeles2005tuning}%
  \BibitemOpen
  \bibfield  {author} {\bibinfo {author} {\bibfnamefont {M.~A.}\ \bibnamefont
  {Serrano}}\ and\ \bibinfo {author} {\bibfnamefont {M.}~\bibnamefont
  {Bogu\~n\'a}},\ }\bibfield  {title} {\bibinfo {title} {Tuning clustering in
  random networks with arbitrary degree distributions},\ }\href
  {https://doi.org/10.1103/PhysRevE.72.036133} {\bibfield  {journal} {\bibinfo
  {journal} {Phys. Rev. E}\ }\textbf {\bibinfo {volume} {72}},\ \bibinfo
  {pages} {036133} (\bibinfo {year} {2005})}\BibitemShut {NoStop}%
\bibitem [{\citenamefont {Toivonen}\ \emph {et~al.}(2006)\citenamefont
  {Toivonen}, \citenamefont {Onnela}, \citenamefont {Saram{\"a}ki},
  \citenamefont {Hyv{\"o}nen},\ and\ \citenamefont
  {Kaski}}]{toivonen2006model}%
  \BibitemOpen
  \bibfield  {author} {\bibinfo {author} {\bibfnamefont {R.}~\bibnamefont
  {Toivonen}}, \bibinfo {author} {\bibfnamefont {J.-P.}\ \bibnamefont
  {Onnela}}, \bibinfo {author} {\bibfnamefont {J.}~\bibnamefont
  {Saram{\"a}ki}}, \bibinfo {author} {\bibfnamefont {J.}~\bibnamefont
  {Hyv{\"o}nen}},\ and\ \bibinfo {author} {\bibfnamefont {K.}~\bibnamefont
  {Kaski}},\ }\bibfield  {title} {\bibinfo {title} {A model for social
  networks},\ }\href {https://doi.org/10.1016/j.physa.2006.03.050} {\bibfield
  {journal} {\bibinfo  {journal} {Physica A}\ }\textbf {\bibinfo {volume}
  {371}},\ \bibinfo {pages} {851} (\bibinfo {year} {2006})}\BibitemShut
  {NoStop}%
\bibitem [{\citenamefont {Bianconi}\ \emph {et~al.}(2014)\citenamefont
  {Bianconi}, \citenamefont {Darst}, \citenamefont {Iacovacci},\ and\
  \citenamefont {Fortunato}}]{bianconi2014triadic}%
  \BibitemOpen
  \bibfield  {author} {\bibinfo {author} {\bibfnamefont {G.}~\bibnamefont
  {Bianconi}}, \bibinfo {author} {\bibfnamefont {R.~K.}\ \bibnamefont {Darst}},
  \bibinfo {author} {\bibfnamefont {J.}~\bibnamefont {Iacovacci}},\ and\
  \bibinfo {author} {\bibfnamefont {S.}~\bibnamefont {Fortunato}},\ }\bibfield
  {title} {\bibinfo {title} {Triadic closure as a basic generating mechanism of
  communities in complex networks},\ }\href
  {https://doi.org/10.1103/PhysRevE.90.042806} {\bibfield  {journal} {\bibinfo
  {journal} {Phys. Rev. E}\ }\textbf {\bibinfo {volume} {90}},\ \bibinfo
  {pages} {042806} (\bibinfo {year} {2014})}\BibitemShut {NoStop}%
\bibitem [{\citenamefont {Newman}(2009)}]{newman2009random}%
  \BibitemOpen
  \bibfield  {author} {\bibinfo {author} {\bibfnamefont {M.~E.~J.}\
  \bibnamefont {Newman}},\ }\bibfield  {title} {\bibinfo {title} {Random graphs
  with clustering},\ }\href {https://doi.org/10.1103/PhysRevLett.103.058701}
  {\bibfield  {journal} {\bibinfo  {journal} {Phys. Rev. Lett.}\ }\textbf
  {\bibinfo {volume} {103}},\ \bibinfo {pages} {058701} (\bibinfo {year}
  {2009})}\BibitemShut {NoStop}%
\bibitem [{\citenamefont {Gleeson}(2009)}]{gleeson2009bond}%
  \BibitemOpen
  \bibfield  {author} {\bibinfo {author} {\bibfnamefont {J.~P.}\ \bibnamefont
  {Gleeson}},\ }\bibfield  {title} {\bibinfo {title} {Bond percolation on a
  class of clustered random networks},\ }\href
  {https://doi.org/10.1103/PhysRevE.80.036107} {\bibfield  {journal} {\bibinfo
  {journal} {Phys. Rev. E}\ }\textbf {\bibinfo {volume} {80}},\ \bibinfo
  {pages} {036107} (\bibinfo {year} {2009})}\BibitemShut {NoStop}%
\bibitem [{\citenamefont {Karrer}\ and\ \citenamefont
  {Newman}(2010)}]{karrer2010random}%
  \BibitemOpen
  \bibfield  {author} {\bibinfo {author} {\bibfnamefont {B.}~\bibnamefont
  {Karrer}}\ and\ \bibinfo {author} {\bibfnamefont {M.~E.~J.}\ \bibnamefont
  {Newman}},\ }\bibfield  {title} {\bibinfo {title} {Random graphs containing
  arbitrary distributions of subgraphs},\ }\href
  {https://doi.org/10.1103/PhysRevE.82.066118} {\bibfield  {journal} {\bibinfo
  {journal} {Phys. Rev. E}\ }\textbf {\bibinfo {volume} {82}},\ \bibinfo
  {pages} {066118} (\bibinfo {year} {2010})}\BibitemShut {NoStop}%
\bibitem [{\citenamefont {Mann}\ \emph {et~al.}(2021)\citenamefont {Mann},
  \citenamefont {Smith}, \citenamefont {Mitchell},\ and\ \citenamefont
  {Dobson}}]{mann2021random}%
  \BibitemOpen
  \bibfield  {author} {\bibinfo {author} {\bibfnamefont {P.}~\bibnamefont
  {Mann}}, \bibinfo {author} {\bibfnamefont {V.~A.}\ \bibnamefont {Smith}},
  \bibinfo {author} {\bibfnamefont {J.~B.~O.}\ \bibnamefont {Mitchell}},\ and\
  \bibinfo {author} {\bibfnamefont {S.}~\bibnamefont {Dobson}},\ }\bibfield
  {title} {\bibinfo {title} {Random graphs with arbitrary clustering and their
  applications},\ }\href {https://doi.org/10.1103/PhysRevE.103.012309}
  {\bibfield  {journal} {\bibinfo  {journal} {Phys. Rev. E}\ }\textbf {\bibinfo
  {volume} {103}},\ \bibinfo {pages} {012309} (\bibinfo {year}
  {2021})}\BibitemShut {NoStop}%
\bibitem [{\citenamefont {Cirigliano}\ \emph {et~al.}(2024)\citenamefont
  {Cirigliano}, \citenamefont {Castellano}, \citenamefont {Baxter},\ and\
  \citenamefont {Tim{\'a}r}}]{cirigliano2024strongly}%
  \BibitemOpen
  \bibfield  {author} {\bibinfo {author} {\bibfnamefont {L.}~\bibnamefont
  {Cirigliano}}, \bibinfo {author} {\bibfnamefont {C.}~\bibnamefont
  {Castellano}}, \bibinfo {author} {\bibfnamefont {G.~J.}\ \bibnamefont
  {Baxter}},\ and\ \bibinfo {author} {\bibfnamefont {G.}~\bibnamefont
  {Tim{\'a}r}},\ }\bibfield  {title} {\bibinfo {title} {Strongly clustered
  random graphs via triadic closure: An exactly solvable model},\ }\href
  {https://doi.org/10.1103/PhysRevE.109.024306} {\bibfield  {journal} {\bibinfo
   {journal} {Physical Review E}\ }\textbf {\bibinfo {volume} {109}},\ \bibinfo
  {pages} {024306} (\bibinfo {year} {2024})}\BibitemShut {NoStop}%
\bibitem [{\citenamefont {Cirigliano}(2025)}]{cirigliano2025how}%
  \BibitemOpen
  \bibfield  {author} {\bibinfo {author} {\bibfnamefont {L.}~\bibnamefont
  {Cirigliano}},\ }\bibfield  {title} {\bibinfo {title} {How universal is the
  mean-field universality class for percolation in complex networks?},\ }\href
  {https://doi.org/10.1209/0295-5075/ae1322} {\bibfield  {journal} {\bibinfo
  {journal} {Europhysics Letters}\ }\textbf {\bibinfo {volume} {152}},\
  \bibinfo {pages} {31002} (\bibinfo {year} {2025})}\BibitemShut {NoStop}%
\bibitem [{\citenamefont {Bhat}\ \emph {et~al.}(2017)\citenamefont {Bhat},
  \citenamefont {Shrestha},\ and\ \citenamefont
  {H\'ebert-Dufresne}}]{bhat2017exotic}%
  \BibitemOpen
  \bibfield  {author} {\bibinfo {author} {\bibfnamefont {U.}~\bibnamefont
  {Bhat}}, \bibinfo {author} {\bibfnamefont {M.}~\bibnamefont {Shrestha}},\
  and\ \bibinfo {author} {\bibfnamefont {L.}~\bibnamefont
  {H\'ebert-Dufresne}},\ }\bibfield  {title} {\bibinfo {title} {Exotic phase
  transitions of $k$-cores in clustered networks},\ }\href
  {https://doi.org/10.1103/PhysRevE.95.012314} {\bibfield  {journal} {\bibinfo
  {journal} {Phys. Rev. E}\ }\textbf {\bibinfo {volume} {95}},\ \bibinfo
  {pages} {012314} (\bibinfo {year} {2017})}\BibitemShut {NoStop}%
\bibitem [{\citenamefont {Flajolet}\ and\ \citenamefont
  {Odlyzko}(1990)}]{flajolet1990singularity}%
  \BibitemOpen
  \bibfield  {author} {\bibinfo {author} {\bibfnamefont {P.}~\bibnamefont
  {Flajolet}}\ and\ \bibinfo {author} {\bibfnamefont {A.}~\bibnamefont
  {Odlyzko}},\ }\bibfield  {title} {\bibinfo {title} {Singularity analysis of
  generating functions},\ }\href {https://doi.org/10.1137/0403019} {\bibfield
  {journal} {\bibinfo  {journal} {SIAM J. Discret. Math.}\ }\textbf {\bibinfo
  {volume} {3}},\ \bibinfo {pages} {216} (\bibinfo {year} {1990})}\BibitemShut
  {NoStop}%
\bibitem [{\citenamefont {Del~Genio}\ \emph {et~al.}(2011)\citenamefont
  {Del~Genio}, \citenamefont {Gross},\ and\ \citenamefont
  {Bassler}}]{delgenio2011all}%
  \BibitemOpen
  \bibfield  {author} {\bibinfo {author} {\bibfnamefont {C.~I.}\ \bibnamefont
  {Del~Genio}}, \bibinfo {author} {\bibfnamefont {T.}~\bibnamefont {Gross}},\
  and\ \bibinfo {author} {\bibfnamefont {K.~E.}\ \bibnamefont {Bassler}},\
  }\bibfield  {title} {\bibinfo {title} {All scale-free networks are sparse},\
  }\href {https://doi.org/10.1103/PhysRevLett.107.178701} {\bibfield  {journal}
  {\bibinfo  {journal} {Phys. Rev. Lett.}\ }\textbf {\bibinfo {volume} {107}},\
  \bibinfo {pages} {178701} (\bibinfo {year} {2011})}\BibitemShut {NoStop}%
\bibitem [{\citenamefont {Baek}\ \emph {et~al.}(2012)\citenamefont {Baek},
  \citenamefont {Kim}, \citenamefont {Ha},\ and\ \citenamefont
  {Jeong}}]{baek2012fundamental}%
  \BibitemOpen
  \bibfield  {author} {\bibinfo {author} {\bibfnamefont {Y.}~\bibnamefont
  {Baek}}, \bibinfo {author} {\bibfnamefont {D.}~\bibnamefont {Kim}}, \bibinfo
  {author} {\bibfnamefont {M.}~\bibnamefont {Ha}},\ and\ \bibinfo {author}
  {\bibfnamefont {H.}~\bibnamefont {Jeong}},\ }\bibfield  {title} {\bibinfo
  {title} {Fundamental structural constraint of random scale-free networks},\
  }\href {https://doi.org/10.1103/PhysRevLett.109.118701} {\bibfield  {journal}
  {\bibinfo  {journal} {Phys. Rev. Lett.}\ }\textbf {\bibinfo {volume} {109}},\
  \bibinfo {pages} {118701} (\bibinfo {year} {2012})}\BibitemShut {NoStop}%
\bibitem [{\citenamefont {Valigi}\ \emph {et~al.}(2025)\citenamefont {Valigi},
  \citenamefont {Serrano}, \citenamefont {Castellano},\ and\ \citenamefont
  {Cirigliano}}]{valigi2025graphicality}%
  \BibitemOpen
  \bibfield  {author} {\bibinfo {author} {\bibfnamefont {P.}~\bibnamefont
  {Valigi}}, \bibinfo {author} {\bibfnamefont {M.}~\bibnamefont {Serrano}},
  \bibinfo {author} {\bibfnamefont {C.}~\bibnamefont {Castellano}},\ and\
  \bibinfo {author} {\bibfnamefont {L.}~\bibnamefont {Cirigliano}},\ }\bibfield
   {title} {\bibinfo {title} {Graphicality of power-law and double power-law
  degree sequences},\ }\bibfield  {journal} {\bibinfo  {journal} {arXiv
  preprint arXiv:2512.24976}\ }\href
  {https://doi.org/10.48550/arXiv.2512.24976} {10.48550/arXiv.2512.24976}
  (\bibinfo {year} {2025})\BibitemShut {NoStop}%
\bibitem [{\citenamefont {Pastor-Satorras}\ \emph {et~al.}(2001)\citenamefont
  {Pastor-Satorras}, \citenamefont {V\'azquez},\ and\ \citenamefont
  {Vespignani}}]{pastor2001dynamical}%
  \BibitemOpen
  \bibfield  {author} {\bibinfo {author} {\bibfnamefont {R.}~\bibnamefont
  {Pastor-Satorras}}, \bibinfo {author} {\bibfnamefont {A.}~\bibnamefont
  {V\'azquez}},\ and\ \bibinfo {author} {\bibfnamefont {A.}~\bibnamefont
  {Vespignani}},\ }\bibfield  {title} {\bibinfo {title} {Dynamical and
  correlation properties of the internet},\ }\href
  {https://doi.org/10.1103/PhysRevLett.87.258701} {\bibfield  {journal}
  {\bibinfo  {journal} {Phys. Rev. Lett.}\ }\textbf {\bibinfo {volume} {87}},\
  \bibinfo {pages} {258701} (\bibinfo {year} {2001})}\BibitemShut {NoStop}%
\bibitem [{\citenamefont {Catanzaro}\ \emph {et~al.}(2005)\citenamefont
  {Catanzaro}, \citenamefont {Bogu\~n\'a},\ and\ \citenamefont
  {Pastor-Satorras}}]{catanzaro2005generation}%
  \BibitemOpen
  \bibfield  {author} {\bibinfo {author} {\bibfnamefont {M.}~\bibnamefont
  {Catanzaro}}, \bibinfo {author} {\bibfnamefont {M.}~\bibnamefont
  {Bogu\~n\'a}},\ and\ \bibinfo {author} {\bibfnamefont {R.}~\bibnamefont
  {Pastor-Satorras}},\ }\bibfield  {title} {\bibinfo {title} {Generation of
  uncorrelated random scale-free networks},\ }\href
  {https://doi.org/10.1103/PhysRevE.71.027103} {\bibfield  {journal} {\bibinfo
  {journal} {Phys. Rev. E}\ }\textbf {\bibinfo {volume} {71}},\ \bibinfo
  {pages} {027103} (\bibinfo {year} {2005})}\BibitemShut {NoStop}%
\bibitem [{\citenamefont {Latora}\ \emph {et~al.}(2017)\citenamefont {Latora},
  \citenamefont {Nicosia},\ and\ \citenamefont {Russo}}]{latora2017complex}%
  \BibitemOpen
  \bibfield  {author} {\bibinfo {author} {\bibfnamefont {V.}~\bibnamefont
  {Latora}}, \bibinfo {author} {\bibfnamefont {V.}~\bibnamefont {Nicosia}},\
  and\ \bibinfo {author} {\bibfnamefont {G.}~\bibnamefont {Russo}},\ }\href
  {http://www.complex-networks.net/programs.html} {\emph {\bibinfo {title}
  {Complex Networks: Principles, Methods and Applications}}}\ (\bibinfo
  {publisher} {Cambridge University Press},\ \bibinfo {year}
  {2017})\BibitemShut {NoStop}%
\bibitem [{\citenamefont {Yin}\ \emph {et~al.}(2019)\citenamefont {Yin},
  \citenamefont {Benson},\ and\ \citenamefont {Leskovec}}]{yin2019local}%
  \BibitemOpen
  \bibfield  {author} {\bibinfo {author} {\bibfnamefont {H.}~\bibnamefont
  {Yin}}, \bibinfo {author} {\bibfnamefont {A.~R.}\ \bibnamefont {Benson}},\
  and\ \bibinfo {author} {\bibfnamefont {J.}~\bibnamefont {Leskovec}},\
  }\bibfield  {title} {\bibinfo {title} {The local closure coefficient: A new
  perspective on network clustering},\ }in\ \href
  {https://doi.org/10.1145/3289600.3290991} {\emph {\bibinfo {booktitle}
  {Proceedings of the Twelfth ACM International Conference on Web Search and
  Data Mining}}},\ \bibinfo {series and number} {WSDM '19}\ (\bibinfo
  {publisher} {Association for Computing Machinery},\ \bibinfo {address} {New
  York, NY, USA},\ \bibinfo {year} {2019})\ p.\ \bibinfo {pages}
  {303–311}\BibitemShut {NoStop}%
\bibitem [{\citenamefont {Feller}(1991)}]{feller1991introduction}%
  \BibitemOpen
  \bibfield  {author} {\bibinfo {author} {\bibfnamefont {W.}~\bibnamefont
  {Feller}},\ }\href
  {https://www.wiley.com/en-us/An+Introduction+to+Probability+Theory+and+Its+Applications%2C+Volume+1%2C+3rd+Edition-p-9780471257080}
  {\emph {\bibinfo {title} {An introduction to probability theory and its
  applications, Volume 1}}},\ Vol.~\bibinfo {volume} {1}\ (\bibinfo
  {publisher} {John Wiley \& Sons},\ \bibinfo {year} {1991})\BibitemShut
  {NoStop}%
\end{thebibliography}%

\end{document}